\documentclass[useAMS,usenatbib]{mn2e}

\usepackage{graphicx}
\usepackage{txfonts}
\usepackage[british]{babel}
\usepackage{latexsym}
\usepackage{amssymb}

\title[Does a prestellar core always become protostellar ?]{Does a prestellar core always become protostellar? Tracing the evolution of cores from the prestellar to protostellar phase}
\author[Anathpindika. S. and  Di Francesco, James]{Anathpindika. S.$^{1}$\thanks{E-mail:
 sumedh$\_$a@iiap.res.in; James.DiFrancesco@nrc-cnrc.gc.ca} and  Di Francesco, James$^{2,3}$\footnotemark[1]\thanks{}\\
$^{1}$Indian Institute of Astrophysics, Bangalore 560034, India\\
$^{2}$ National Research Council Canada, Herzberg Institute of Astrophysics, 5071 West Saanich Road, Victoria, BC, V9E 2E7, Canada; \\$^{3}$ Department of Astronomy \& Physics, University of Victoria, P.O. Box 355, STN CSC, Victoria, BC, V8W 3P6, Canada}
\begin{document}

\date{Accepted 0000 December 00. Received 0000 December 00; in original form 1988 October 11}

\pagerange{\pageref{firstpage}--\pageref{lastpage}} \pubyear{2002}

\maketitle

\label{firstpage}

\begin{abstract}
Recently, a subset of starless cores whose thermal Jeans mass is apparently overwhelmed by the mass of the core has been identified, e.g., the core {\small L183}. In literature, massive cores such as this one are often referred to as "super-Jeans cores". As starless cores are perhaps on the cusp of forming stars, a study of their dynamics will improve our understanding of the transition from the prestellar to the protostellar phase. In the present work we use non-magnetic polytropes  belonging originally to the family of the Isothermal sphere. For the purpose, perturbations were applied to individual polytropes, first by replacing the isothermal gas with a gas that was cold near the centre of the polytrope and relatively warm in the outer regions, and second, through a slight compression of the polytrope by raising the external confining pressure. Using this latter configuration we identify thermodynamic conditions under which a core is likely to remain starless. In fact, we also argue that the attribute "super-Jeans" is subjective and that these cores do not formally violate the Jeans stability criterion. On the basis of our test results we suggest that gas temperature in a star-forming cloud is crucial towards the formation and evolution of a core. Simulations in this work were performed using the particle-based Smoothed Particle Hydrodynamics algorithm. However, to establish numerical convergence of the results we suggest similar tests with a grid-scheme, such as the Adaptive mesh refinement. 
\end{abstract}


\section{Introduction}
Stars form out of relatively dense volumes of molecular gas, typically denser than $\sim 10^{4}$ cm$^{-3}$, generally referred to as cores.  The dynamical stability of a core is determined by the relative importance of thermal  (which includes turbulence), and magnetic pressure against its own gravity.  A ``starless'' core is one in which there is no evidence of star-formation. A gravitationally bound starless core with internal temperatures typically of order a few Kelvin, and a flat central density relative to the protostellar core, is called a pre-protostellar or simply prestellar core (Ward-Thompson \emph{et al.} 1994); see also Di Francesco et al. (2007) for a relatively recent review of the observed characteristics of starless and prestellar cores. 

Having become sufficiently massive  by accreting gas from the ambient medium, a core may evolve to become centrally dense, mimicking the density profile of the singular isothermal sphere. The protostellar phase  of a core is thus one in which a subsequent collapse leads to the formation of a young stellar object, often detected due to its emission at infrared wavelengths from warm dust surrounding the young protostar. Young stars also launch outflows and therefore, are useful tracers of star-formation activity.

The process through which a core becomes protostellar is not fully understood, though a possible evolutionary track for prestellar cores was proposed by Simpson \emph{et al.} (2011). This process, however, may be crucial towards explaining the distribution of stellar masses characterised by the initial mass function (e.g., see Ward-Thompson \emph{et. al.} 2007). In general, prestellar cores should collapse when internal support such as thermal or magnetic pressure is overwhelmed by gravity. Several cores that appear to violate this simple evolutionary model have been identified, however.  For instance, the core {\small L694-2}, though starless, shows signs of infall (e.g., Harvey \emph{et al.} 2003), many others show signs of outward motions or both inward and outward motions, like {\small B68} (Lada et al. 2003). 

In addition, several cores have been identified as being starless despite having either: (i) masses apparently greater than their thermal Jeans mass, for a non-magnetised core (e.g., see Sadavoy et al. 2010a,b), or (ii) relatively low internal velocity dispersion, of order $\sim$0.1 km/s to 0.3 km/s with cold interiors (e.g., cores {\small L1498} and {\small L1517B}; Tafalla \emph{et al.} 2004). Cores of type (i) are also referred to as ``super-Jeans'' in literature. Including line observations to determine velocity dispersion, Schnee et al. (2012) recently determined that the continuum-derived masses of a sub-sample of five super-Jeans cores in Ophiuchus and Perseus are about equal to the virial mass of a critically stable Bonnor-Ebert sphere, suggesting weakly that the super-Jeans cores are bound even when non-thermal motions are considered. The super-Jeans nature of these starless cores is of course debatable and will be examined in this contribution.

Observed lifetimes (and estimated long lifetimes of the order of several free-fall times), of some prestellar cores suggest they are not in free-fall, and perhaps supported by some mechanism. Magnetic fields and turbulence may play a key role in this regard. For instance, cores located in magnetised regions of a cloud should evolve on the ambipolar diffusion timescale, usually of the order of a few Myrs, although cores close to super-criticality are likely to evolve much faster, on the order of only a Myr (e.g., Ciolek \& Basu 2001). The age of a core can be estimated through a study of the gas-phase abundances of molecular species such as  CS, NH$_{3}$, CO and its isotopologues, and other complex molecules. Ages so derived for the cores {\small L1498} and {\small L1517B} (Tafalla \emph{et al.} 2004), of the order of 1-2 Myrs are indeed, consistent with the ambipolar timescale for weakly magnetised cores. Similarly, the more chemically evolved cores in the Pipe Nebula lie in strongly magnetised parts of the cloud, and probably will also evolve on the classical ambipolar diffusion timescale, of order ten free fall times (Frau \emph{et al.} 2010). According to numerical simulations, however, turbulence, the other possible support against gravity,  decays on a relatively short timescale, typically less than a sound-crossing time (e.g., Pavlovski \emph{et al.} 2002). Consequently, it is unlikely to explain the long lifetimes of some prestellar cores. These cores have therefore  been, the subject of intensive investigation by  several authors, e.g., Evans \emph{et al.} (2001), Crapsi \emph{et al.} (2005a,b), Hatchell \emph{et al.} (2007), Ward-Thompson \emph{et al.} (2010), to name only a few.

Some authors have studied the formation of cores numerically via the growth of density perturbations. V{\' a}zquez-Semadeni \emph{et al.} (2005) reported that some of these perturbations condensed out as small clumps, a few of which collapsed while others re-expanded before eventually dispersing. Similar conclusions were also derived by Klessen \emph{et al.}(2005) from their simulations. Galv{\'a}n-Madrid \emph{et al.}(2007), simulating turbulent, magnetically super-critical clouds, derived results consistent with the findings of the latter authors. In fact, cores forming in the simulations by Galv{\'a}n-Madrid \emph{et al.}(2007) evolved on a relatively short timescale, typically 1-2 Myrs, as suggested for the cores L1498 and L1517B. The well-known starless core {\small B68} on the other hand is probably in a critical equilibrium and might perform low-amplitude oscillations  when perturbed weakly (Keto \emph{et al.} 2006; Redman \emph{et al.} 2006). In the present study, we will examine the stability of cores and determine new conditions that are likely to retard them from collapse. The plan of the paper is as follows. We commence in \S 2 by briefly discussing the problem, followed by a short recap of the physical properties of the Isothermal sphere and an introduction to the numerical scheme used here. Simulations are discussed in \S 3 and \S 4 before concluding in \S 5.

\begin{table*}
 \centering
 \begin{minipage}{140mm}
  \caption{List of models tested.}
  \begin{tabular}{@{}lllll@{}}
  \hline
   Serial & Confinement & Polytropic & Temperature & Observation\\
   No.    &   & model & distribution ($T(r)$) & \\
 \hline
   1 & U\footnote{U-Unconfined polytrope} & A\footnote{Originally a singular Isothermal sphere but now superposed with a radially dependent temperature profile, $T(r)$} &  Equation 7\footnote{$T(r)$ defined by Equation (7)}, $T_{in}$ = 6 K, $T_{out}$ = 15 K & Collapse\\
   2 & C\footnote{C-confinement by external pressure defined by Eqn. (3)} & A &  Equation 7, $T_{in}$ = 6 K, $T_{out}$ = 15 K, $T_{ICM}$ = 50 K & Collapse \\
   3 & C & B\footnote{Originally a Bonnor-Ebert sphere but now superposed with a radially dependent temperature profile, $T(r)$}& Equation 7, $T_{in}$ = 6 K, $T_{out}$ = 15 K, $T_{ICM}$ = 20 K & Collapse \\
   4 & C & B & Equation 7, $T_{in}$ = 10 K, $T_{out}$ = 35 K, $T_{ICM}$ = 40 K & Oscillations \\
   5 & C & B & Equation 7, $T_{in}$ = 6 K, $T_{out}$ = 35 K, $T_{ICM}$ = 40 K  & Oscillations \\
   6 & C & I\footnote{Isothermal}  & $T_{0}$ = 20 K, $T_{ICM}$ = 25 K & Collapse \\
\hline
\end{tabular}
\end{minipage}
\end{table*}

\begin{figure}
  \vspace*{0pt}
  \includegraphics[angle=0,width=8.cm]{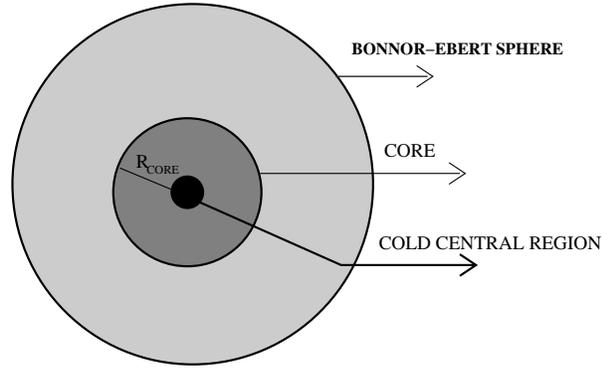}
  \caption{A schematic illustrating the nomenclature adopted in the rest of this paper. Shown here is a core of radius, $r_{core}$, assembled by radial inflow of gas. The core has been shaded grey and encloses the cold central region shaded black, while the outer annulus in a lighter shade represents the polytrope, a Bonnor-Ebert sphere (a variety of the Isothermal sphere used in some simulations here). }
\end{figure}

\section[]{Initial conditions}
\subsection[]{Description of the problem}
The possibility of the formation of a prestellar core out of a natal volume of gas will be examined in this work. This hypothesis will be examined by running a set of comparative models that will use two varieties of the Isothermal sphere. Note that, "Isothermal" with "I" will be reserved for the polytrope generated by the solution to the Lane-Emden equation, Eqn. (2) below; "i", will be used to denote uniform gas-temperature. The isothermal gas within the respective polytropes will be superposed by a more realistic temperature distribution, often found in prestellar cores and defined by Eqn. (7) below. The resulting assembly is demonstrated by the schematic sketched in Fig. 1, where $r_{core}$ is the radius of the core assembled at the centre of the agitated polytrope. A core in this work is defined as the pocket of gas within the model volume having average density greater than a threshold, $\rho_{thresh}= 10^{-19}$ g cm$^{-3}$$\ \sim 10^{4}$ cm$^{-3}$. The polytrope is perturbed by inwardly propagating disturbance triggered due to a slight increase in the external pressure, achieved by suitably raising the temperature of the gas confining the polytrope.

\subsection[]{The Isothermal sphere revisited}
The principal equation characterising a polytrope is the well-known Lane-Emden equation,
\begin{equation}
\frac{1}{\xi^{2}}\frac{d\mu}{d\xi} + \psi^{n} = 0.
\end{equation}
Gas pressure, $P$, under the isothermal approximation, is related to its density, $\rho$, via a simple equation of state of the form, $P = a_{0}^{2}\rho^{\gamma}$, where  $a_{0}=\Big(\frac{k_{B}T_{0}}{\bar{m}}\Big)^{1/2}$, is the sound-speed for isothermal gas held at temperature, $T_{0}$. The exponent, $\gamma = 1 + 1/n$, where the polytropic index, $n=\infty$, for an isothermal gas and the mean molecular mass, $\bar{m}=4\times 10^{-24}$ g, for a typical cosmic mixture (i.e., a mixture of molecular hydrogen and helium).

The Isothermal sphere belongs to a larger family of polytropes defined by Eqn. (1). Under the isothermal approximation Eqn. (1) reduces to
\begin{equation}
\frac{d\mu}{d\xi} = \xi^{2}e^{-\psi};
\end{equation}
where $\mu = \Big(\xi^{2}\frac{d\psi}{d\xi}\Big)$, the Lane-Emden function, $\psi = \Big(\frac{\rho}{\rho_{c}}\Big)^{1/n}$, and $\xi$ is the dimensionless radius (Chandrasekar 1939). Stability characteristics of the Isothermal sphere show that it is  inherently unstable against gravity for any radius, $\xi_{B}\equiv\xi>\xi_{crit}\equiv 6.45$.  The density, $\rho(\xi)$, of the Isothermal sphere decreases radially outward and at $\xi=\xi_{crit}$, it falls to about 70\% of its value at the centre. 

Configurations with $\xi\leq\xi_{crit}$, called a Bonnor-Ebert (BE) sphere, is confined by a finite external pressure, $P_{ext}$, and supported internally by thermal pressure against gravity\footnote{The free-fall time, $t_{ff}=\Big(\frac{3\pi}{32G\bar{\rho}}\Big)^{1/2}$, defined for gas with average density, $\bar{\rho}$, though strictly inapplicable to a polytrope, has been adopted  as the unit of time in the rest of this article for mere convenience.}. The pressure confining a BE sphere is
\begin{equation}
P_{ext} = \rho_{c}e^{-\psi(\xi_{B})}a_{0}^{2},
\end{equation}
where $\rho_{c}$ is the central density, and $\psi(\xi)$, the asymptotic solution to the Isothermal Lane-Emden equation (Bonnor 1956, Ebert 1955). By demanding pressure equilibrium at the boundary of the Bonnor-Ebert sphere the minimum temperature, $T_{0}$, necessary to maintain stability can be readily obtained by combining Eqn. (3) with the equation of state for an isothermal gas. Thus,
\begin{equation}
T_{0}\equiv T(\xi_{B}) = \frac{\bar{m}}{k_{B}}\frac{\rho_{c}}{\rho(\xi_{B})}e^{-\psi(\xi_{B})};
\end{equation}
 $k_{B}$ is the Boltzmann constant. Finally, the maximum mass within a {\small BE} sphere of radius $\xi_{B}$ is,
\begin{equation}
M_{BE}\equiv M(\xi=\xi_{B}) = 4\pi\rho_{c}R_{0}^{3}\mu(\xi_{B}),
\end{equation}
where the length scaling factor, $R_{0} = \Big(\frac{a_{0}^{2}}{4\pi G\rho_{c}}\Big)^{1/2}$, and $\mu(\xi)$ has been defined above for Eqn. (2). 

\begin{figure*}
\vbox to 70mm{\vfil 
\mbox{\includegraphics[angle=270,width=8.cm]{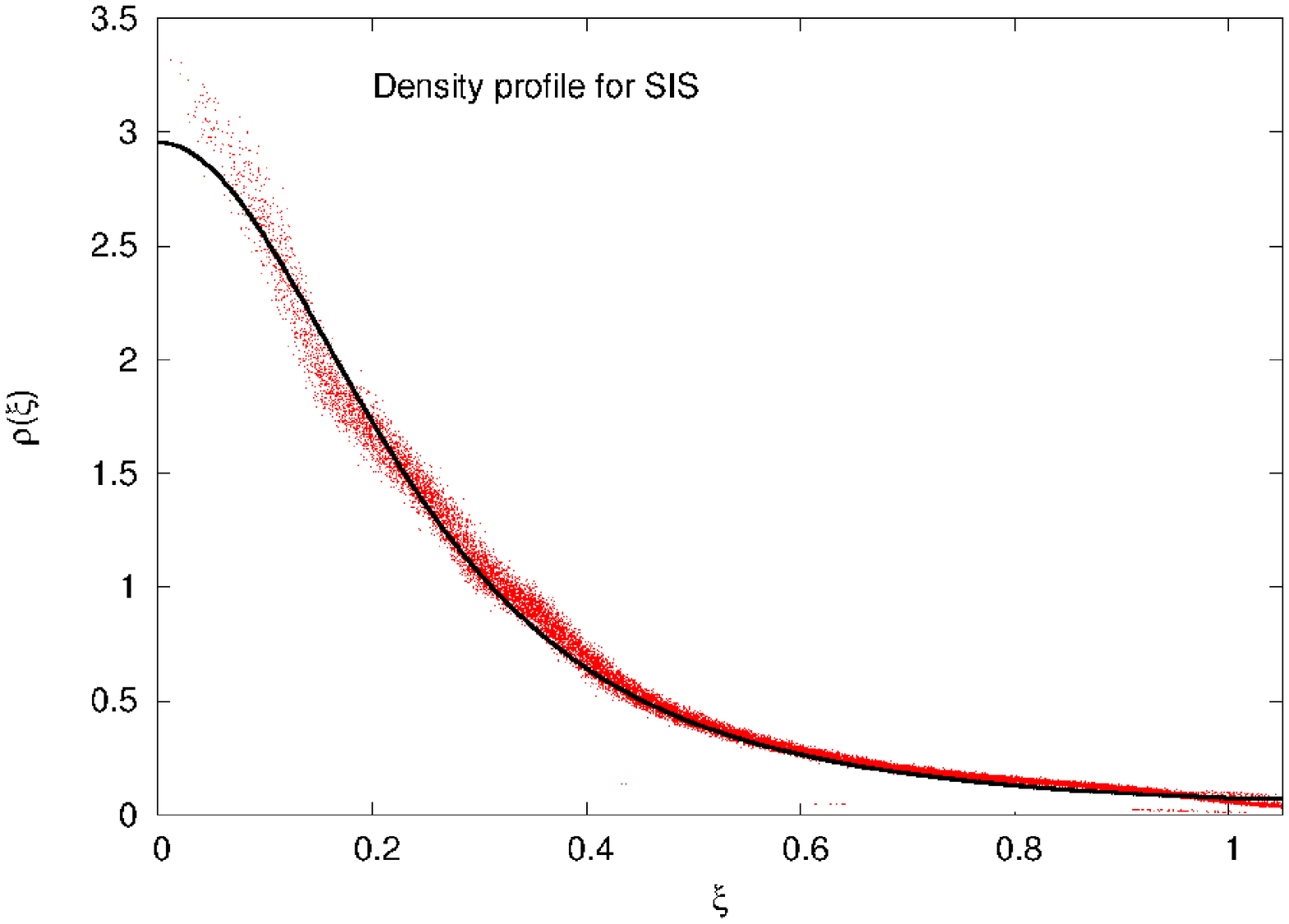}
       \includegraphics[angle=270,width=8.cm]{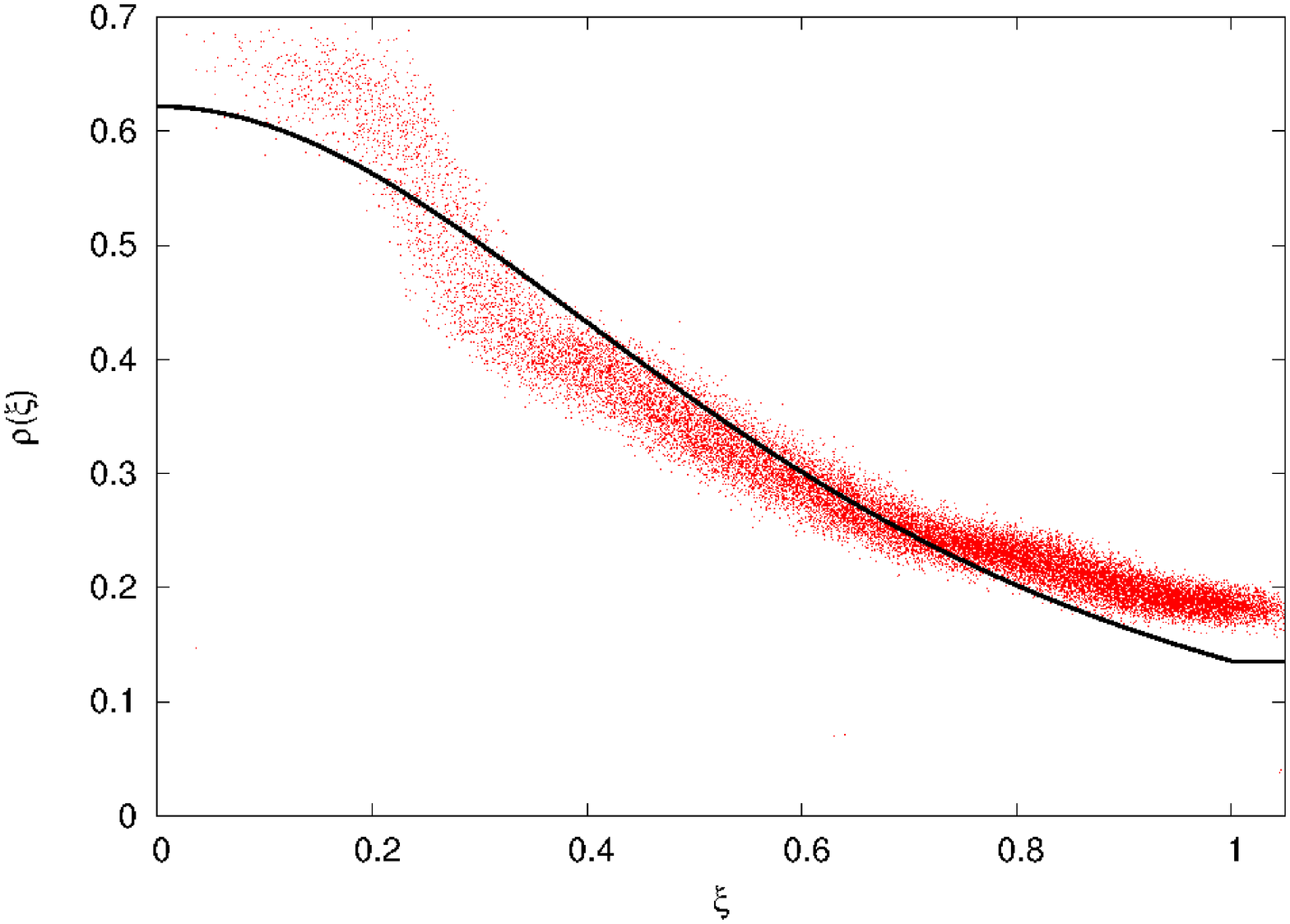}}
 \caption{Shown in the left- and right-panel is the respective radial density profile for a singular Isothermal sphere ($\xi_{B}$=10) and a Bonnor-Ebert sphere($\xi_{B}$=4), each of unit mass and radius; the plot is made in dimensionless units. Marked by dots is the {\small SPH} calculation of density while result from analytic calculation is plotted by a continuous line.}
   \vfil} \label{landfig}
\end{figure*}

\subsection[]{Smoothed Particle Hydrodynamics ({\small SPH})}
The well-known Lagrangian, particle-based scheme {\small SPH} finds a wide-range of applications in computational hydrodynamics, including astrophysical hydrodynamics (e.g., Monaghan 2005). {\small SPH} particles are not point masses in the strictest sense so that each particle has finite spatial extent defined by its smoothing length, $h_{i}$, apart from other state variables, viz. mass, velocity and temperature. Normal  {\small SPH} particles exert the gravitational and the thermal force on each other and forces are calculated at every time step. After-ward, the position of each particle is revised by integrating the equation of motion. {\small SPH} particles interact with each other through a parametrised numerical viscosity that prevents them from penetrating each other. Viscous dissipation becomes effective for pairs of particles approaching each other and diminishes rapidly for those receding.

Computational efficiency is acquired by stacking particles on an octal tree, like the one suggested by Barnes \& Hut (1986), where leaf cells at the bottom of the tree contain less than four particles. Calculation of the net force and the search for nearest neighbours, $N_{neibs}$, of a particle are carried out with this tree. Though contributions from distant cells are approximated, the accuracy of calculations is improved by including the quadrupole moments. The time integration is performed by using the multiple time-stepping algorithm where each particle is assigned an appropriate time step. 

The temperature of gas particles is calculated using the radiative transfer scheme introduced by Stamatellos \emph{et al.} (2007). The scheme proceeds by constructing a polytrope typically of index, $n$=2, around each {\small SPH} particle, and then calculates the average optical depth for that particle.
The optical depth determines the extent to which a particle is shielded and calculates the corresponding change in internal energy of that particle. The algorithm uses parametrised opacity, calculated such that it accounts for physical processes that dominate a density-temperature range. See Table 1 in Stamatellos \emph{et. al.} (2007) for a list of processes covered. Finally the revised temperature of that particle is calculated by solving the equation for thermal equilibrium. Gas is assumed to be primarily composed of hydrogen(70\%) and helium(30 \%), even as the scheme successfully mimics the relevant gas-phase chemistry within a typical prestellar core. The numerical code {\small SEREN} used in this work is a well-tested algorithm; see Hubber \emph{et al.} (2011) for further details.  

An {\small SPH} particle denser than a predefined threshold, $\rho_{sink}\sim 10^{-15}$ g cm$^{-3}$, is replaced by a sink if it also satisfies two other qualifications : (a) it has the desired number of neighbours, at least $N_{neibs}$, and (b) it is gravitationally bounded  (Bate \emph{et al.} 1995).  The sink particle in the present set of simulations represents effectively the protostellar phase of a putative star-forming core. The reason for choosing a relatively low $\rho_{sink}$ will become apparent below. This choice of $\rho_{sink}$ also implies that temperature of the  gas within our test polytropes exhibits little change throughout a simulation as adiabatic heating is ineffective below $\sim 10^{-12}$ g cm$^{-3}$. Simulations discussed here were performed on a 12-thread dual-Xeon X5675 processor of the Hydra super-computing cluster
 at the Indian Institute of Astrophysics. \\ \\
\textbf{\emph{Resolution}}  Simulations were performed with $5.5\times 10^{6}$ particles, of which the number of gas particles, $N_{gas}= 5\times 10^{6}$, while the rest were particles representing the intercloud medium ({\small ICM}). The initial average {\small SPH} smoothing length, $h_{avg}$, that provides an estimate of the spatial resolution is defined as
\begin{equation}  
h_{avg} = \frac{R_{init}}{2}\Big(\frac{N_{neibs}}{N_{gas}}\Big)^{1/3},
\end{equation}
where $N_{neibs}$=50. Thus, $h_{avg}\sim 1.5\times 10^{-3}$ pc, so that, $\mathcal{X}\equiv r_{0}/h_{avg}\sim 15$, i.e., the central region within radius, $r_{0}$, is resolved by about 10 {\small SPH} particles; $r_{0}$ is defined by Eqn. 7 below.  Increasing $\mathcal{X}$ by an order of magnitude, would require $N_{gas}$ to be at least 3 orders of magnitude higher than the current choice, well beyond our present computational capabilities. The current choice of $N_{gas}$, as will be seen below, is sufficient, for our extant purpose since we are not keen to resolve details of the protostellar phase.

The minimum resolvable density, $\rho_{res}\sim 7.7\times 10^{-20}$$(N_{gas})^{2}$ g cm$^{-3}$, which for the  parameters  chosen in the present work is $\sim 10^{-10}$ g cm$^{-3}$,  about five orders of magnitude greater than $\rho_{sink}$. Similarly, the  minimum resolvable mass, $M_{res}$, is
\begin{displaymath}
M_{res} = 2N_{neibs}\Big(\frac{M_{init}}{N_{gas}}\Big)\equiv 10^{-4} \mathrm{M}_{\odot},
\end{displaymath}
where the bracketed quantity in the expression above is simply the mass of a single {\small SPH} particle, $m_{i}$ (Bate \& Burkert 1997). The Jeans mass, $M_{J}$, at the sink-formation threshold of $\rho_{sink}\sim 10^{-15}$ g cm$^{-3}$ and temperature 10 K (see also Table 1), is $\sim 1.6\times 10^{-2}$ M$_{\odot}$. Thus, $\frac{M_{J}}{M_{res}}\sim$ 1500, implying, a Jeans mass will be resolved by about 1500 particles. Our simulations are therefore well resolved to study the onset of the protostellar phase, and can even be extended to the formation of the first core. However, the present study does not proceed beyond the protostellar phase.

\subsection[]{The test polytropes}
Out of the infinitely large family of the Isothermal sphere, we initially opt for one each with radius on either side of its critical radius, $\xi_{crit}$, with $\xi_{B}$ = 4 and $\xi_{B}$ = 10, respectively. The former is a Bonnor-Ebert sphere while the latter, a singular Isothermal sphere that is inherently unstable to gravity. The initial density profiles for the respective configurations are shown in Figure 2. Details about assembling the respective polytropes have been described in the following subsection, \S 2.5. 

Having assembled each of the two settled polytropes, a non-isothermal temperature distribution, $T(r)$, of the type
\begin{equation}
T(r) = T_{out} - \frac{(T_{out}-T_{in})}{1 + (r/r_{0})^{1.5}},
\end{equation}
 initially suggested by Crapsi \emph{et al.} (2007) for the starless core {\small L1544} was overlaid.  Here $T_{out}\equiv T(r>r_{0})$, $T_{in}\equiv T(r<r_{0})$, and $r_{0}\sim$ 5000 AU. This expression accounts for the relatively low temperatures within the interiors of prestellar cores, which according to detailed radiative transfer modeling, seldom exceed 10 K. For at typical prestellar densities the gas-dust coupling becomes dominant and dust particles cool efficiently via thermal radiation even as gas-phase coolants freeze out onto cold dust grains (e.g., Goldsmith 2001). The external regions though, remain relatively warm due to a combination of a number of heating sources like cosmic rays, energetic external photons, turbulence, etc. While this is the case for real cores, in the present work however, there is no external source of heating.

 The initial mass, $M_{init}$, and radius, $r_{init}$, of the test polytrope in the present work is respectively, 5.5 M$_{\odot}$ and 0.15 pc. To maintain stability for the above-defined choices of mass and radius, gas within the test polytropes will have to be maintained at $T_{0}\gtrsim$ 51 K, according to Eqn. (4) . However, our choice of the gas temperature renders the polytropes dynamically unstable; see Table 1.

\begin{figure*}
  \hspace{1.cm}
  \begin{minipage}{\linewidth}
  \centering
  \mbox{\includegraphics[angle=270,width=8.5cm]{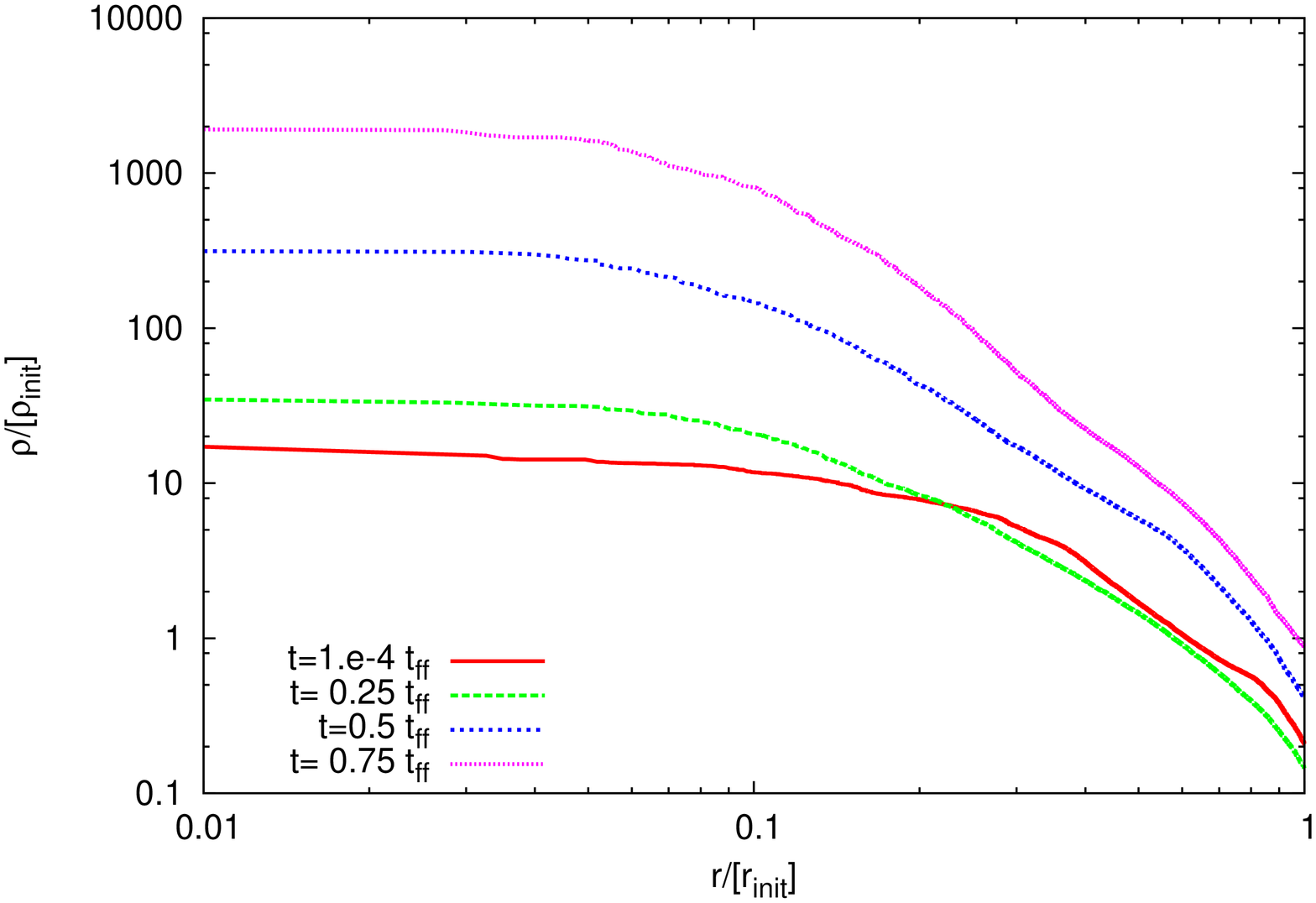}
        \includegraphics[angle=270,width=8.5cm]{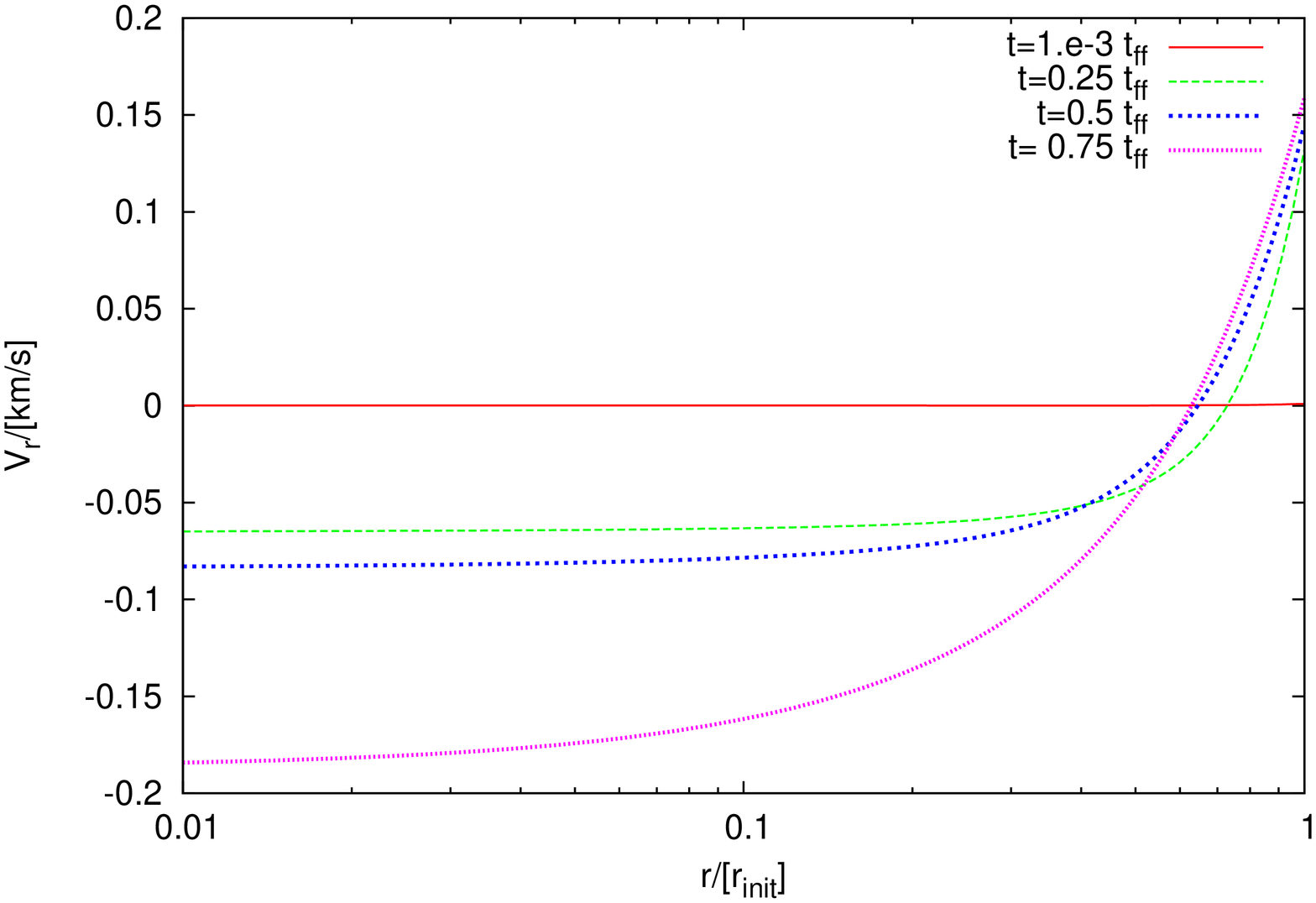}}
  \caption{The coeval plots in the left- and right-hand panel show respectively the radial profile of the gas-density, and the radial component of gas-velocity within the polytrope A for case 1. Radial distance is marked in units of the precollapse radius, $r_{init}$, of the polytrope. Evidently, as the interiors of the polytrope collapse, the gas starting from rest falls-in increasingly rapidly although, it always remains subsonic (sound speed within the envelope of the precollapse polytrope is 0.23 km/s).}
  \end{minipage}
\end{figure*}

\begin{figure}
  \vspace*{24pt}
  \centering
  \includegraphics[angle=270,width=8.cm]{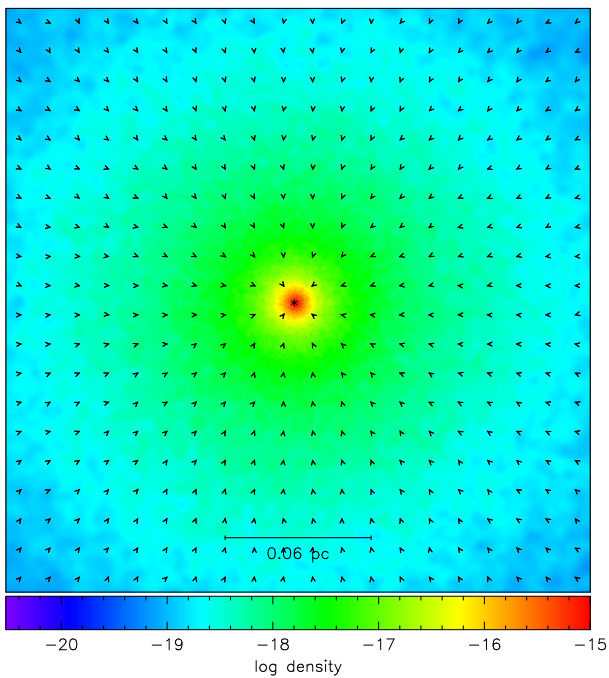}
  \caption{A rendered density plot showing the mid-plane close to the centre of the polytrope in case 1 at a latter stage(t $\sim$ 0.3 Myr $\equiv$1.1 $t_{ff}$). Overlaid on this plot is the local velocity field shown by arrows, while the sink particle representing a young protostar has been marked by an $\ast$ at the centre of the singular polytrope. The radially inward moving gas within the polytrope can be readily inferred from the direction of the velocity vectors. }
\end{figure} 

\subsection[]{Assembling the polytropes}
Polytropes used in the actual simulations were generated by first assembling two Isothermal spheres of unit mass and radius, one each for $\xi_{B}$= 10 and 4, respectively. We remind, $\xi_{B}$, is the dimensionless radius of the Isothermal sphere. Polytropes of either type were assembled by positioning particles randomly within them after which they were settled, before finally scaling them to the desired dimensions. A settled polytrope is one in which the Poisson noise associated with the random assembly particles has been significantly dissipated. This was achieved by repeatedly revising particle positions within the polytrope in the presence of artificial viscosity that dissipates thermal noise. The polytrope was prevented from diffusing into vacuum by jacketing it with a warm intercloud medium (ICM), represented by a special type of {\small SPH} particle that only exerts hydrodynamic force on gas particles. The polytrope-{\small ICM} system was then confined by another thin shell of boundary particles, meant to prevent {\small SPH} particles from escaping into vacuum. These boundary particles remain fixed in space and do not exert any force on other {\small SPH} particles.

The number density of particles was adjusted so that pressure equilibrium was maintained across the gas-{\small ICM} interface. Figure 2 shows the density profiles for settled polytropes  of unit dimensions. Having so settled the two polytropes, they were then rescaled to the desired dimensions, and a radially dependent temperature profile defined by Eqn. (7) above was then superposed on either of them. Thus, the polytropes used in actual simulations, with the exception of case 6 listed in Table 1, were no longer isothermal. In the rest of this article the polytrope with radius $\xi_{B}$ = 10, originally a singular Isothermal sphere, but now a singular non-Isothermal sphere, will be referred to as polytrope A. Similarly, the Bonnor-Ebert sphere with $\xi_{B}$ = 4, also no longer Isothermal, will be referred to as polytrope B. Before proceeding to discuss the actual simulations it would be useful to demonstrate the suitability of the {\small SPH} algorithm for the extant purpose. The fact that the algorithm can reproduce the analytically predicted behaviour of the Isothermal sphere of critical radius, $\xi_{B}=\xi_{crit}$, is demonstrated in appendix A below.

\begin{figure}
  \vspace*{5pt}
  \includegraphics[angle=270,width=8.cm]{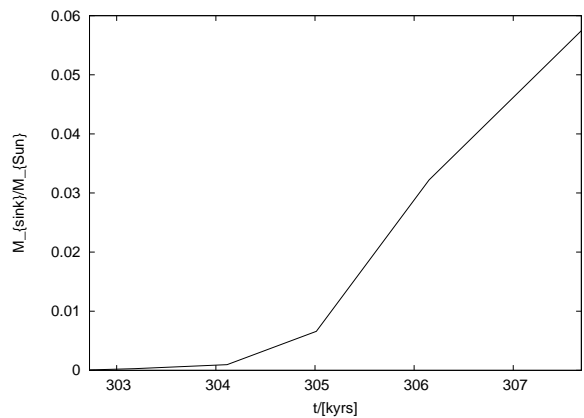}
  \caption{The plot shows the early accretion history of the young protostar in case 1, represented by a sink particle in the simulation. }
\end{figure}

\begin{figure*}
  \hspace{1.cm}
  \begin{minipage}{\linewidth}
  \centering
  \mbox{\includegraphics[angle=270,width=8.5cm]{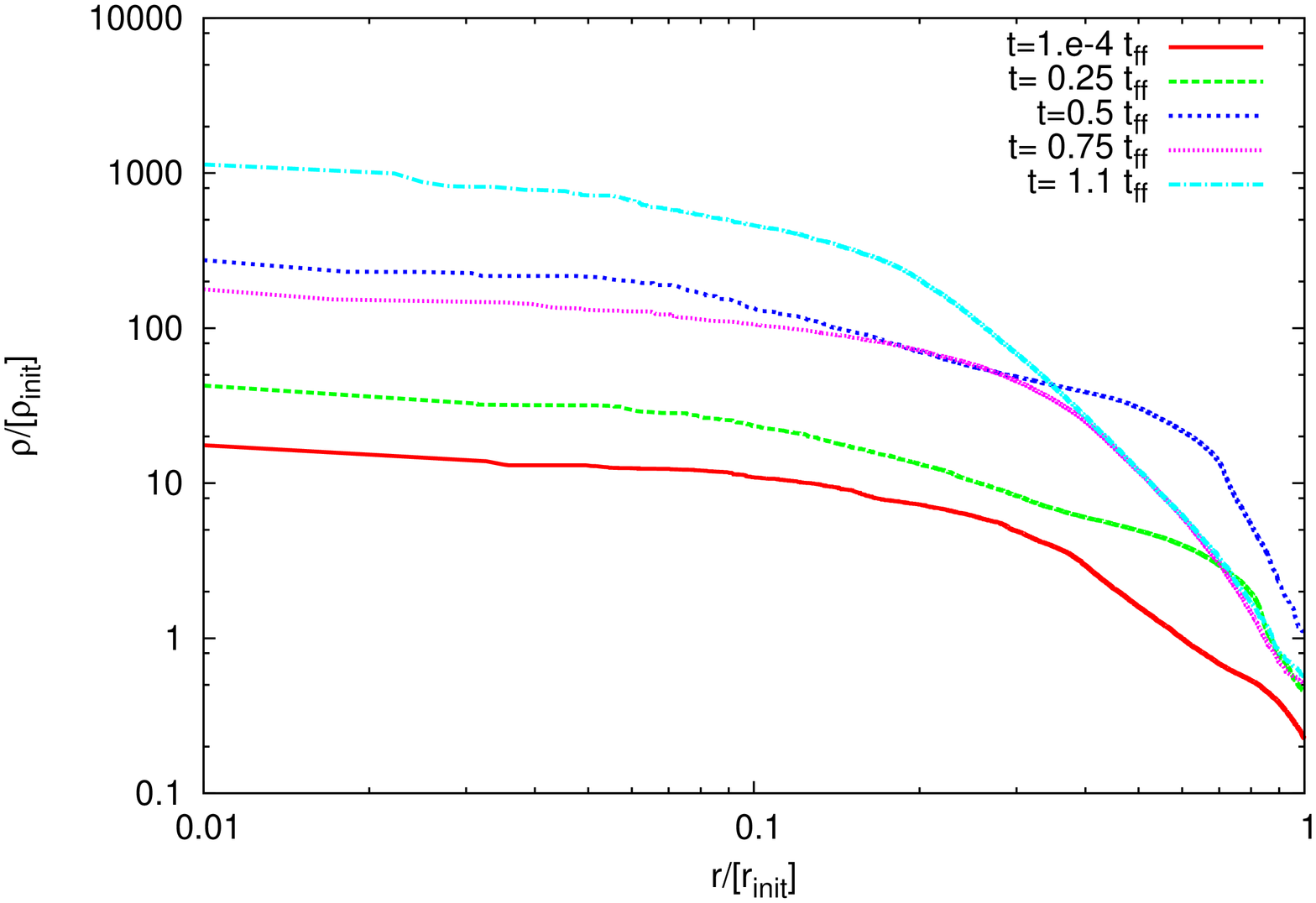}
        \includegraphics[angle=270,width=8.5cm]{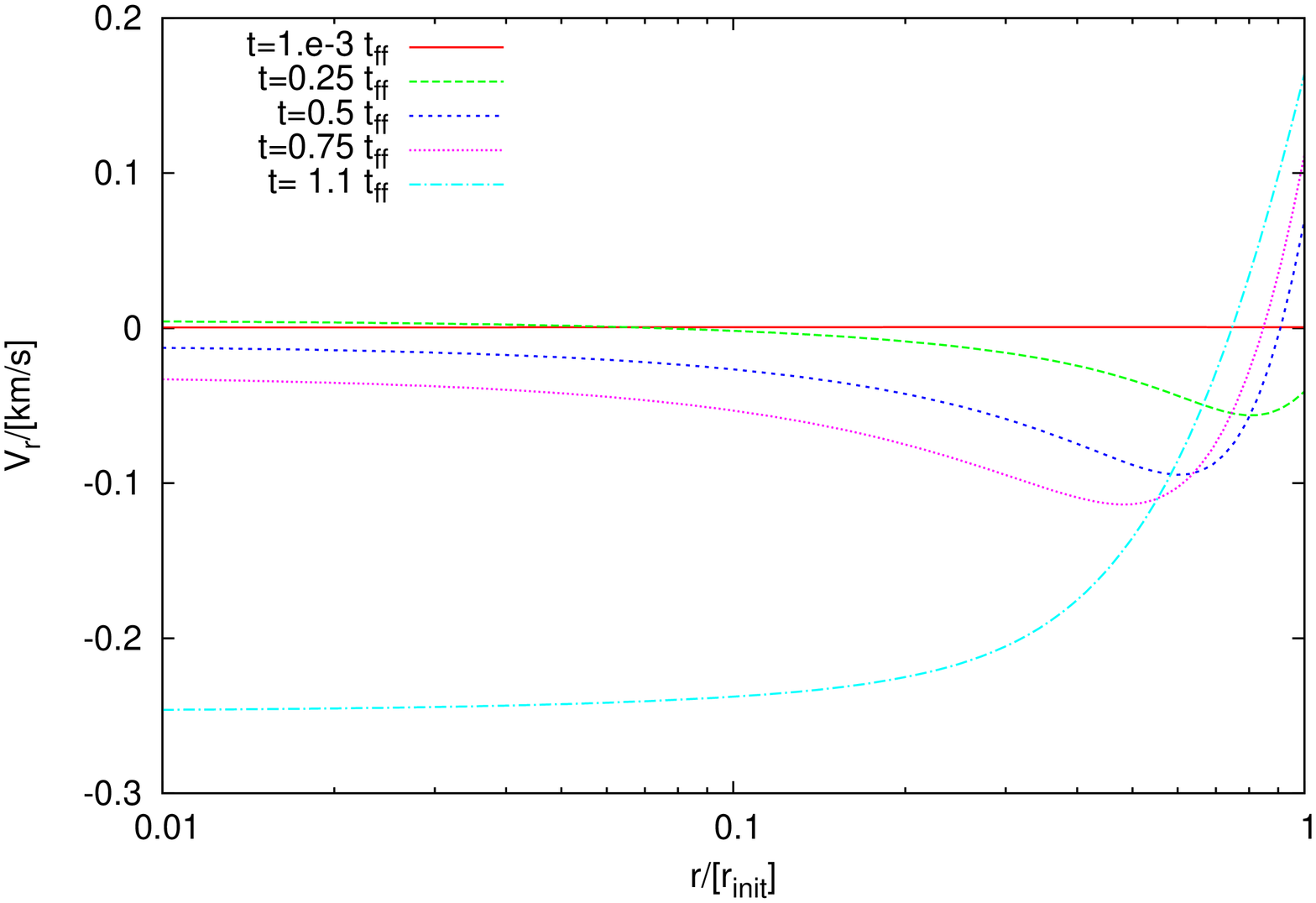}}
  \caption{As in Fig.3, radial profiles of the gas-density and radial velocity within the polytrope for case 2 have been shown respectively in the left-, and right-hand panels. While the plots are generally identical to those in Fig. 3 for case 1, gas close to the centre becomes mildly supersonic as indicated by the blue velocity curve ($t\sim 1.1\ t_{ff}$).}
  \end{minipage}
\end{figure*}

\section[]{Numerical simulations}
We begin by introducing the convention adopted for velocity plots in the rest of this paper. Shown in these plots is the radial velocity of gas within a polytrope at different epochs. Gas moving towards the centre of a polytrope has negative velocity while that moving away from the centre has positive velocity. The direction of fluid motion within a polytrope is also reflected by the direction of velocity vectors overlaid on top of rendered density plots whenever necessary. 

\subsection{Cases 1 and 2 with polytrope A}
The singular polytrope A in cases 1 and 2 when perturbed, promptly collapses, a  result that is hardly surprising.  The polytrope for case 1 is the same as one in vacuum at infinitely large temperature. Thus, while gas within the interiors of this polytrope ( $r/r_{init}\gtrsim 0.6\sim 0.1$ pc), is soon overwhelmed by gravity and begins to collapse, that  in the outer regions diffuses. Starting from rest, the collapsing gas acquires a velocity $\sim 0.2$ km/s. Shown respectively in the left-, and right-hand panels of Fig. 3 is the gas-density and the velocity within the collapsing polytrope at different epochs for this case.

\begin{figure*}
  \hspace{1.cm}
  \begin{minipage}{\linewidth}
  \centering 
   \mbox{\includegraphics[angle=270,width=8.5cm]{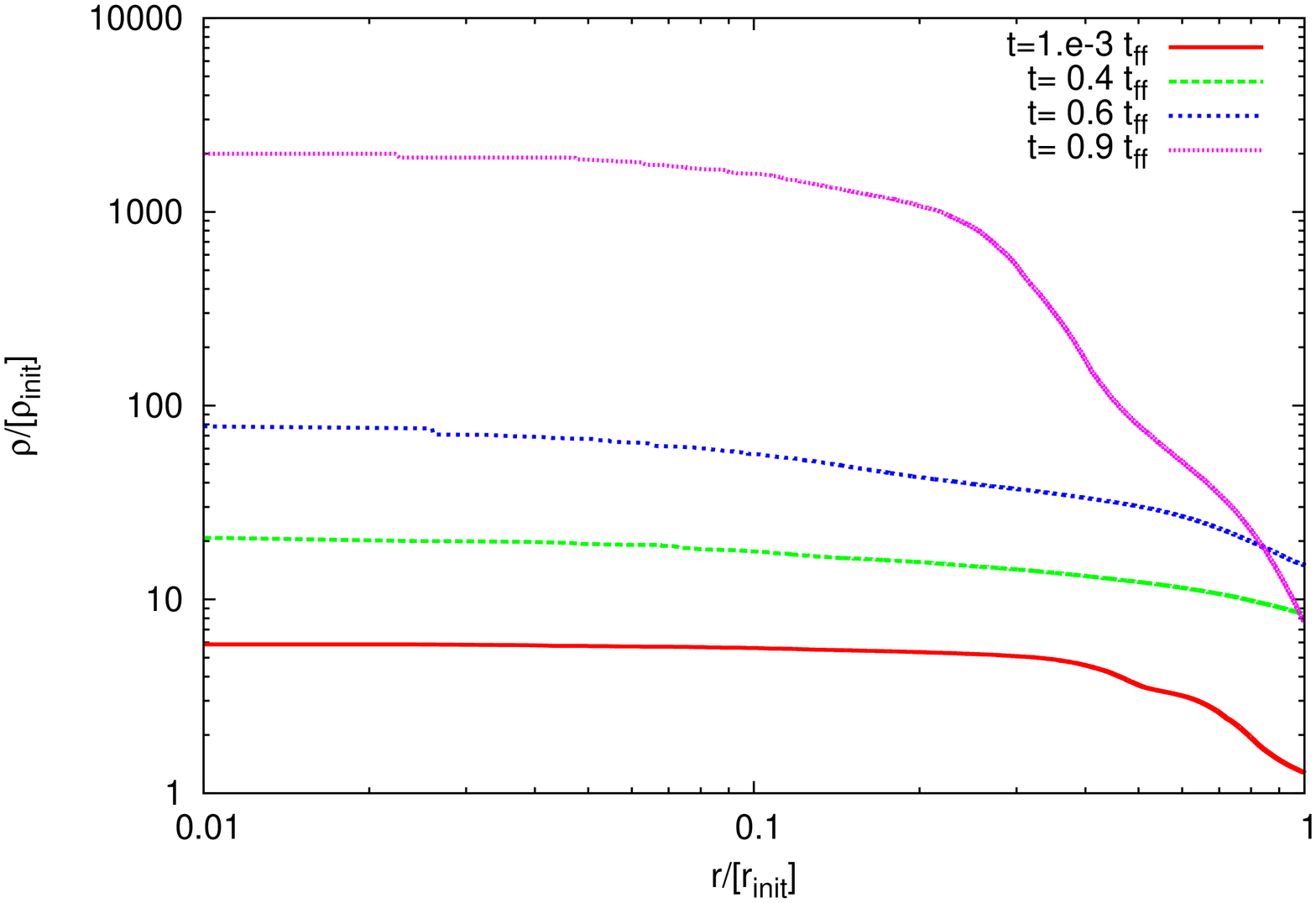}
        \includegraphics[angle=270,width=8.5cm]{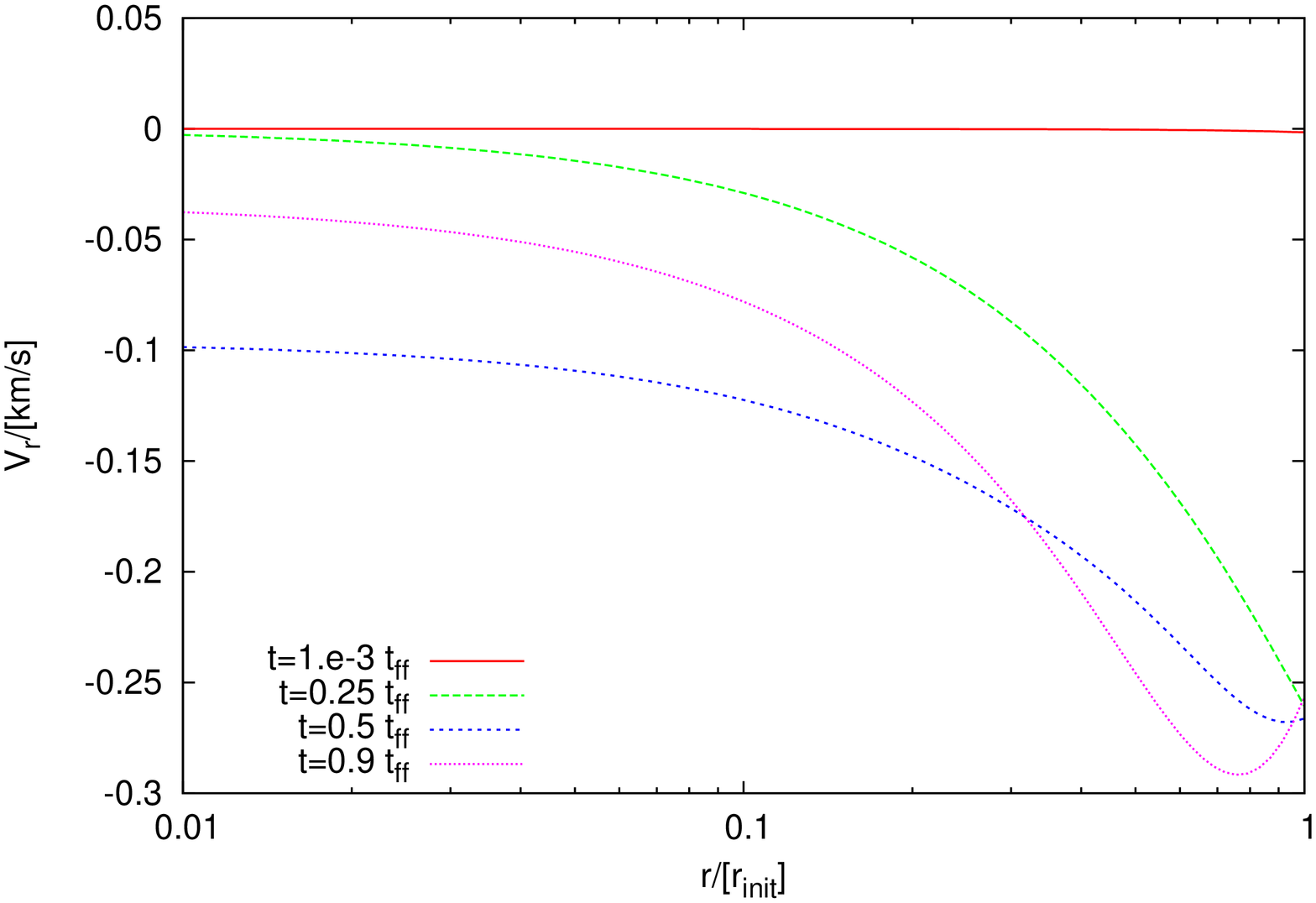}}
  \caption{As in Figure 6, radial profiles of the gas-density and the radial component of gas-velocity within the polytrope for case 3 have been shown in the left- and right-hand panels, respectively. Polytrope B in this case is dynamically unstable and gas close to its centre, starting from rest, is soon overwhelmed by the inwardly propagating, subsonic compression-wave as can be seen from the velocity plots (sound-speed within the polytrope is $\sim$0.23 km/s).}
 \end{minipage}
\end{figure*}

\begin{figure*}
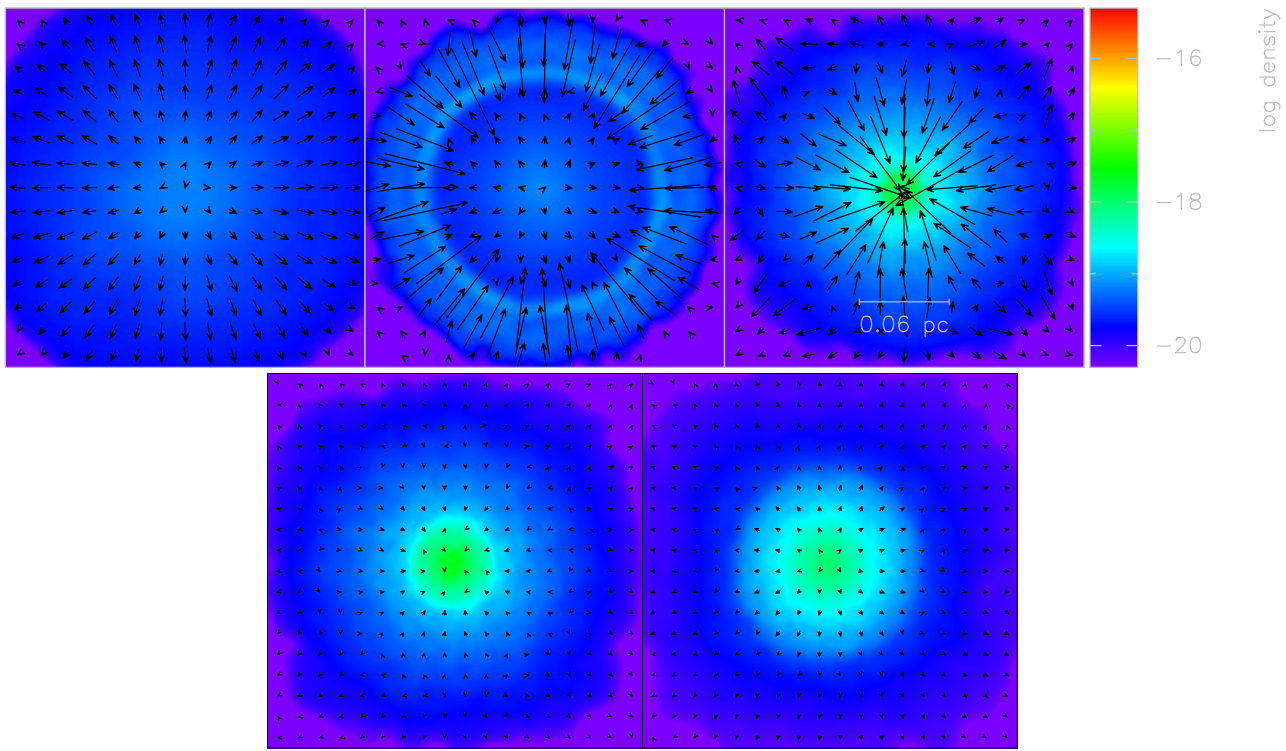

\centering
  \vbox to 115mm{\vfil
  \includegraphics[angle=270,width=17.cm]{C4BEmontg.eps}
  \includegraphics[angle=270,width=10.cm]{C4BEmontg2.eps}
 \caption{\emph{Upper-panel}: A cross-sectional rendered density montage (in units g cm$^{-3}$) showing the first half-cycle of oscillation, where gas assisted by the compression-wave moves inward. As in Fig. 4, the local velocity field is indicated by arrows overlaid on the plots; $t=0.1 t_{ff}, 0.5 t_{ff}, 1 t_{ff}$, respectively for plots in left-, central-, and right-hand panels (1$t_{ff}\sim$ 0.4 Myr). \emph{Lower-panel}: The compression soon makes way for rarefaction, evident from the direction of velocity vectors in the plots on the left- ($t=1 t_{ff}$) and the right-hand panel ($t=1.2 t_{ff}$), respectively. These plots demonstrate the  rebounce of the pressure-supported core. } \vfil}
\end{figure*}

Collapse of the polytrope effectively marks the beginning of the protostellar phase.  Infall is also evident from the overlaid arrows that represent the direction of the local velocity field. The rendered density plot in Fig. 4 shows the mid-plane of the singular polytrope for case 1 where the young protostellar object that forms at the centre of the polytrope is marked by an $\ast$. Gas is  steadily accreted by the young protostar and its mass, $M_{\ast}$, grew almost linearly with time at a rate of a few times 10$^{-5}$ M$_{\odot}$ yr$^{-1}$, as can be seen from its early accretion history shown in Fig. 5. A comparison with the analytic estimate of protostellar accretion-rate, $\dot{M}_{\ast}$, would be useful, and the simplest approximation to which is the Bondi-accretion model. If  $\rho_{s}$ and $a$ are respectively, the average density of the gas being accreted and sound-speed in this gas then, $\dot{M}_{\ast} = 4\pi\rho_{s}r_{s}^{2}a$; $r_{s}=\frac{G M_{init}}{2a^{2}}$, is the sonic radius.  Plugging in the appropriate values yields, $\dot{M}_{\ast} \sim 3\times 10^{-5}$ M$_{\odot}$ yr$^{-1}$, which agrees with that deduced from the early accretion history.  This rate of accretion is also consistent with those reported for typical protostellar cores. Andr{\' e} \emph{at al.} (2007), for instance,  have derived similar values for cores in the Ophiuchus star-forming cloud.  

 The same polytrope as in case 1, but now confined by a finite external pressure defined by Equation (3), was used  in case 2. By contrast, the polytrope in this case is in global collapse that occurs over a free-fall time, slightly longer than that observed in case 1. The density and velocity plots for this case have been shown in Fig. 6.  As in the previous case, gas close to the centre of the polytrope in this case also starts from rest and moves inward, though after a free-fall time the gas-velocity becomes slightly supersonic (blue curve, $t\sim 1.1 t_{ff}$); see Fig. 6.

\subsection{Remaining cases with polytrope B}
We will now discuss the next  set of our simulations, listed as 3, 4 and 5 in Table 1. Reminding that polytrope B is a pressure-confined {\small BE} sphere superposed with a non-isothermal temperature distribution defined by Eqn. (7) would be useful. Results from this set of simulations are mutually contrasting, so while the polytrope in case 3 became singular, it exhibited large-scale radial oscillations under different initial conditions in cases 4 and 5. Inferences drawn from this set of simulations will then be used in \S 4 to hypothesise a scenario in which prestellar cores could be arrested in their evolution to the protostellar phase. As in the first two cases, gas initially at rest within the polytrope in case 3 gradually moves towards the centre of the polytrope as it is swept by  an inwardly propagating compression-wave. The wave though, remains largely sub-sonic even as infalling gas close to the centre gathers momentum. This can be readily seen from the velocity plots on the right-hand panel of Fig. 7.

 The dynamically unstable polytrope in cases 4 and 5 failed to become singular and instead, performed large-scale radial oscillations. As in the previous cases, the inwardly propagating compression-wave was again triggered through a slight increase in the confining pressure, $P_{ext}$. Consequently, the polytrope showed intermittent phases of enhanced central density that eventually exceeded the threshold, $\rho_{thresh}$, for a prestellar core. Rendered density plots overlaid with the local velocity vectors in the upper-panel of Fig. 8 show the compression-phase of the agitated polytrope, while the rarefaction is shown in the lower panel. These plots spanning a little over a free-fall time of the original polytrope readily demonstrate the assembly of a prestellar core due to the radial inflow of gas, assisted by the compression-wave (upper-panel, central-plate),  followed by the first rebounce of the core (lower-panel, right-hand plate). Although the simulation was allowed to continue over six free-fall times, for want of space only a sequence over the first free-fall time has been shown in Fig. 8 here.

\begin{figure}
  \vspace{5pt}
   \includegraphics[width=6.5cm, angle=270]{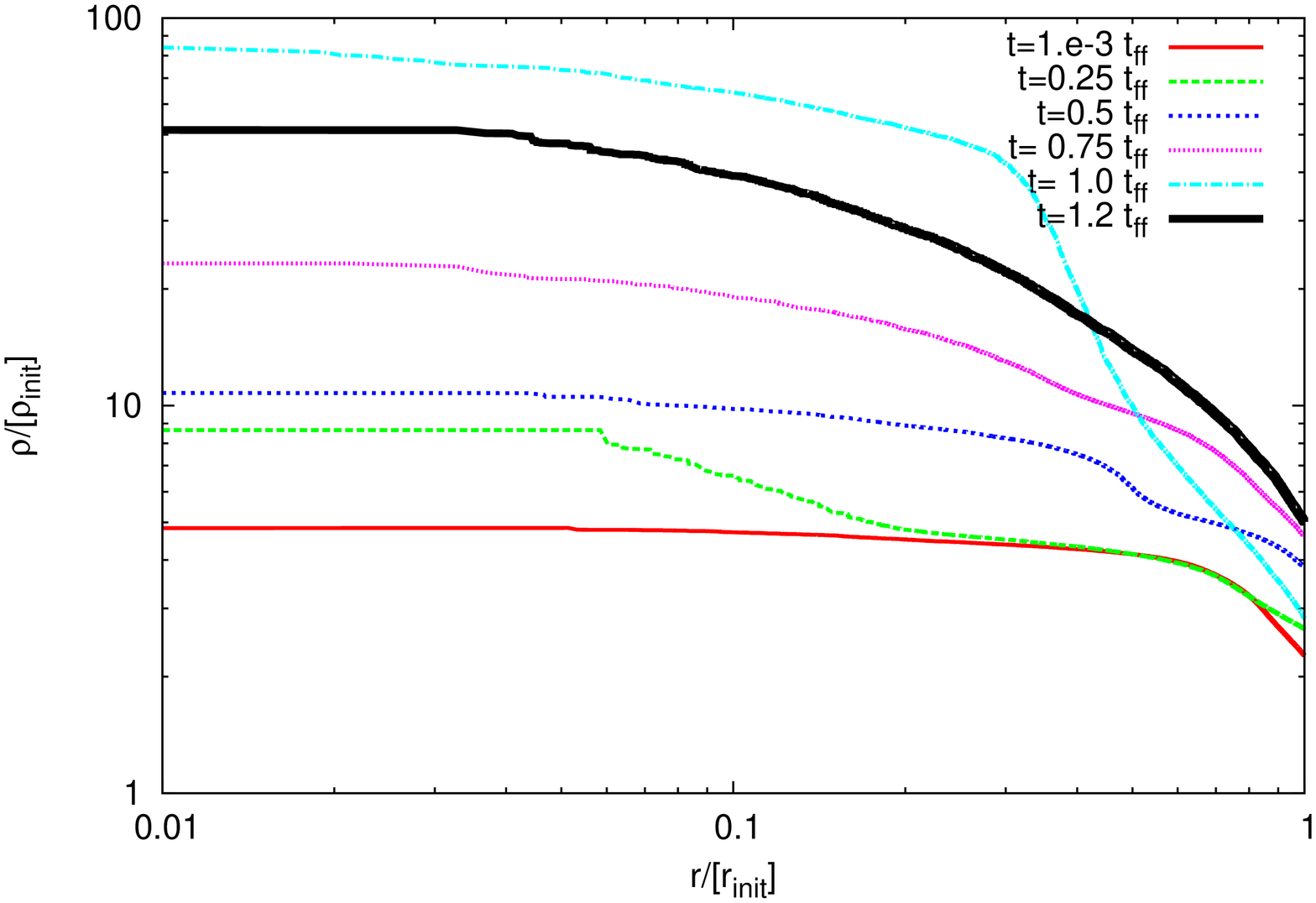}
   \includegraphics[width=6.5cm, angle=270]{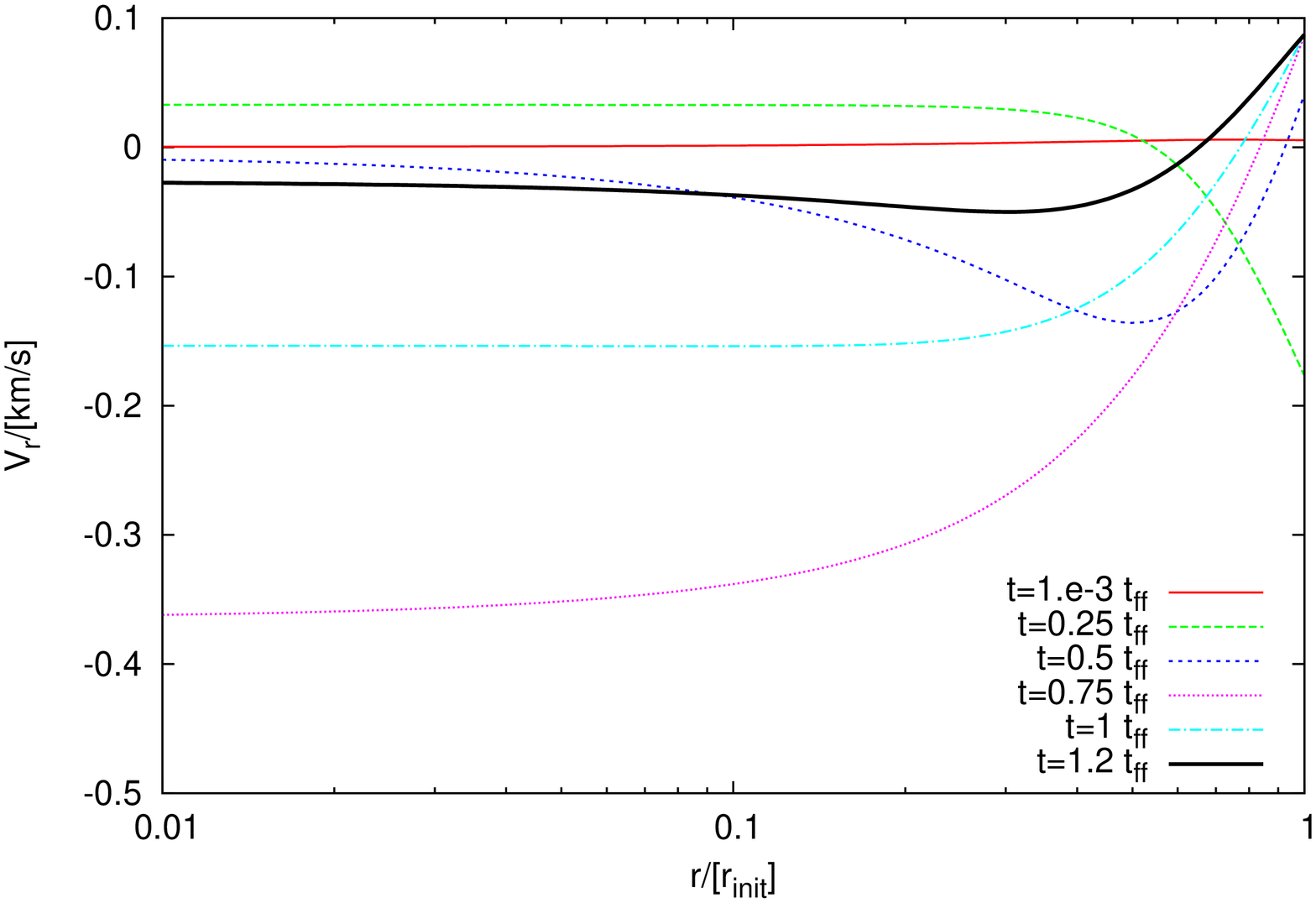}
   \includegraphics[width=6.5cm, angle=270]{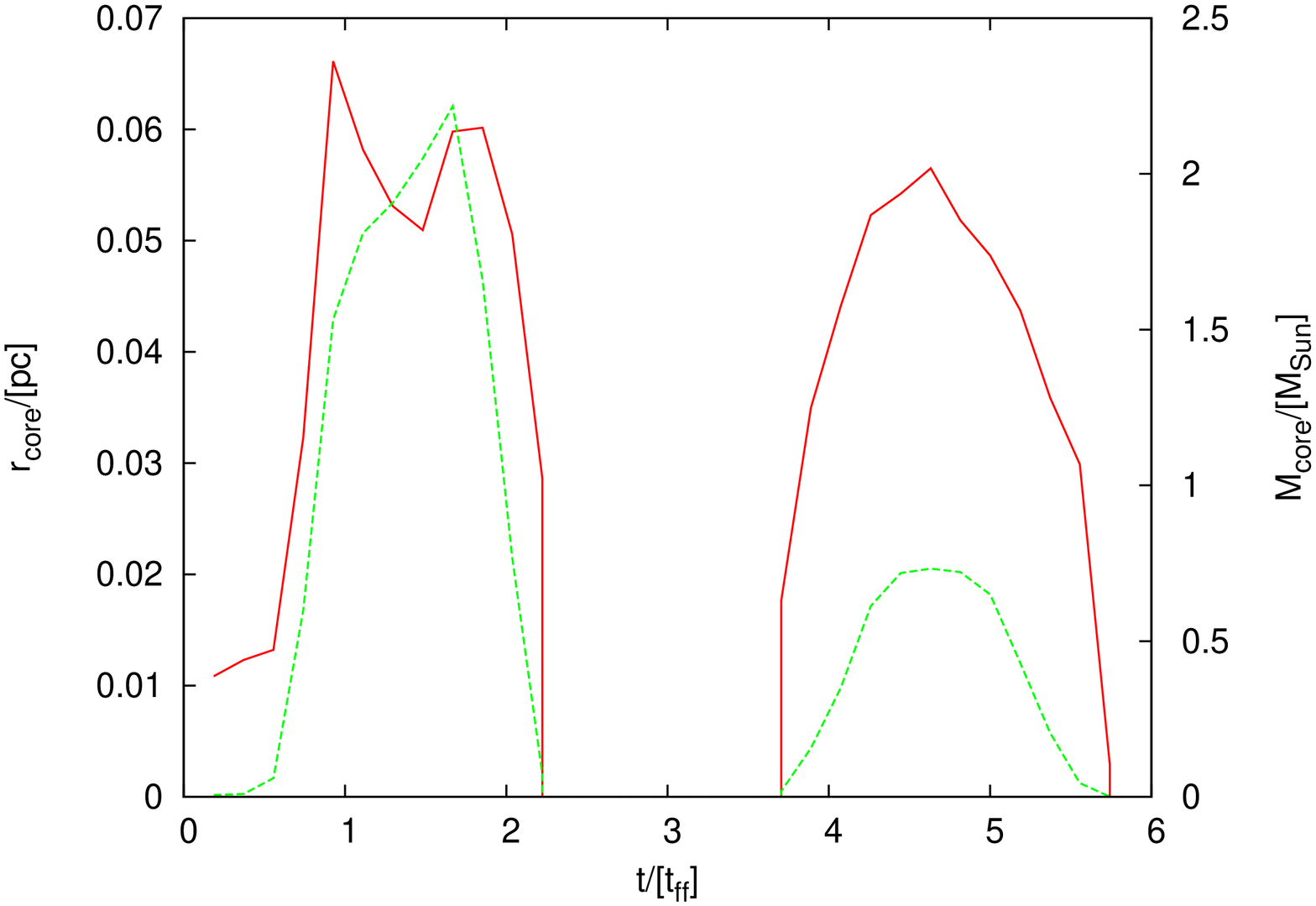}
   \caption{Shown on a logarithmic-scale in the upper-panel is the radial distribution of gas density within the {\small BES} for case 4 at different epochs. A coeval plot showing the radial component of velocity has been shown on the central-panel. The inward flow of gas assisted by the compression-wave is reflected by the negative gas-velocity. The compression-wave, as in the earlier three cases, remains largely sub-sonic, though gas close to the centre briefly becomes transonic (curve for $t\sim 0.75 t_{ff}$). Thereafter the core soon rebounds and gas moves outward as demonstrated by the characteristics for gas density and radial-velocity at $t=1 t_{ff} \& 1.2 t_{ff}$; see also text. The plots on the lower panel show the radius (red curve), and mass (green curve) of the assembled core. }
   \end{figure}

The rendered plots in Fig. 8 are further supplemented by those in Fig. 9 where  radial variation of the gas-density and velocity within the agitated polytrope has been plotted for one oscillation, on the upper and central-panel, respectively. The gas-density within the polytrope steadily rises as the compression-wave moves inward, then having reached a maximum and as a rarefaction sets in, the density decreases ($t\sim 1.2 t_{ff}$). Correspondingly, gas within the polytrope over-run by the inwardly moving compression flows towards the centre with ever-increasing velocity, as demonstrated by the curves corresponding to $t=0.5 t_{ff}$ and $t=0.75 t_{ff}$. Then as  a rarefaction sets-in, gas within the core assembled during the compression-phase begins to move outward gradually and its velocity becomes less negative ($t=1 t_{ff}$). The outwardly directed gas motion associated with the core-rebound is more explicitly visible from the velocity-curve for $t=1.2 t_{ff}$.

Plotted in the lowest panel of Fig. 9 is the mass of the core, $M_{core}$, and its radius, $r_{core}$. The core assembled during the first half cycle of contraction and dissolved in the following rarefaction, is reassembled during the next half-cycle of contraction followed by rarefaction again. This sequence over six free-fall times of the original polytrope is demonstrated by this plot.  The radius of this core  varies typically between 0.01 pc and 0.07 pc, comparable to core-radii reported in several surveys of dense cores (e.g., see Jijina \emph{et al.} 1999, Kirk \emph{et al.} 2007, Belloche \emph{et al.} 2011). In dimensionless units the core radius is, $\xi_{B}^{core}\equiv \frac{r_{core}}{R_{0}}$, where $R_{0}$ is the physical scale-length introduced previously in \S 2.2. Interestingly, $\xi_{B}^{core}$, for the assembled core, as defined above, is 2, confirming its pressure-supported nature. Its rebounce is therefore hardly surprising. That particles exceeding the density threshold, $\rho_{thresh}\sim 10^{-19}$ g cm$^{-3}$, formulate the core  would serve as a useful reminder for our readers at this point.

The critical Bonnor-Ebert mass for the core, $M_{BE}^{core}$, is obtained by simply rewriting Eqn. (5), now with $\xi_{B}^{core}=2$. This yields,
\begin{equation}
M_{BE}^{core}\equiv M_{BE}(\xi_{B}^{core}=2) \sim 0.54\Big(\frac{a_{0}^{2}}{G}\Big)^{3/2}\frac{1}{\rho_{c}^{1/2}}.
\end{equation}
The central density, $\rho_{c}$, and temperature of gas within the core is $\sim 10^{-17}$ g cm$^{-3}$ and 10 K, respectively so that $M_{BE}^{core}\sim$ 0.032 M$_{\odot}$. \textbf{Now the mass enclosed within the radius, $r_{core}=\xi_{B}^{core}R_{0}\sim$ 0.004 pc, is simply, $M_{core}(\xi_{B}^{core})\sim 4\pi\rho_{c} r_{core}^{3}/3\sim$ 0.04 M$_{\odot}$.} \textbf{Thus, $\mathcal{J}\equiv \frac{M_{core}(\xi_{B}^{core})}{M_{BE}^{core}}$, is slightly in excess of unity. 
And further, it is significantly higher than unity when calculated using the mass within the entire core of radius, $\sim 0.01$ pc, pointing to its proclivity to collapse under gravity.} Thus the plot in the lower-panel of Fig. 9 demonstrates the epochs when $\mathcal{J}$ exceeded unity, i.e., portions of the green-curve exceeding $M_{BE}^{core}$ calculated by Eqn. (8) above. Crucially though, the core did not become singular. An important caveat in the above argument is the assumption of a uniform temperature in Eqn. (8) to calculate the critical BE mass, $M_{BE}^{core}$, for the core. This is inappropriate in view of the non-isothermal nature of gas within the test polytrope. In the strictest sense therefore, the stability criterion must be tested at different radii within the polytrope. This is done by calculating the Jeans mass, $M_{J}(r)$, and the BE-mass, $M_{BE}(r)$, as a function of the core radius.

\begin{figure*}
 \vspace{5pt}
  \includegraphics[width=12cm, angle=270.]{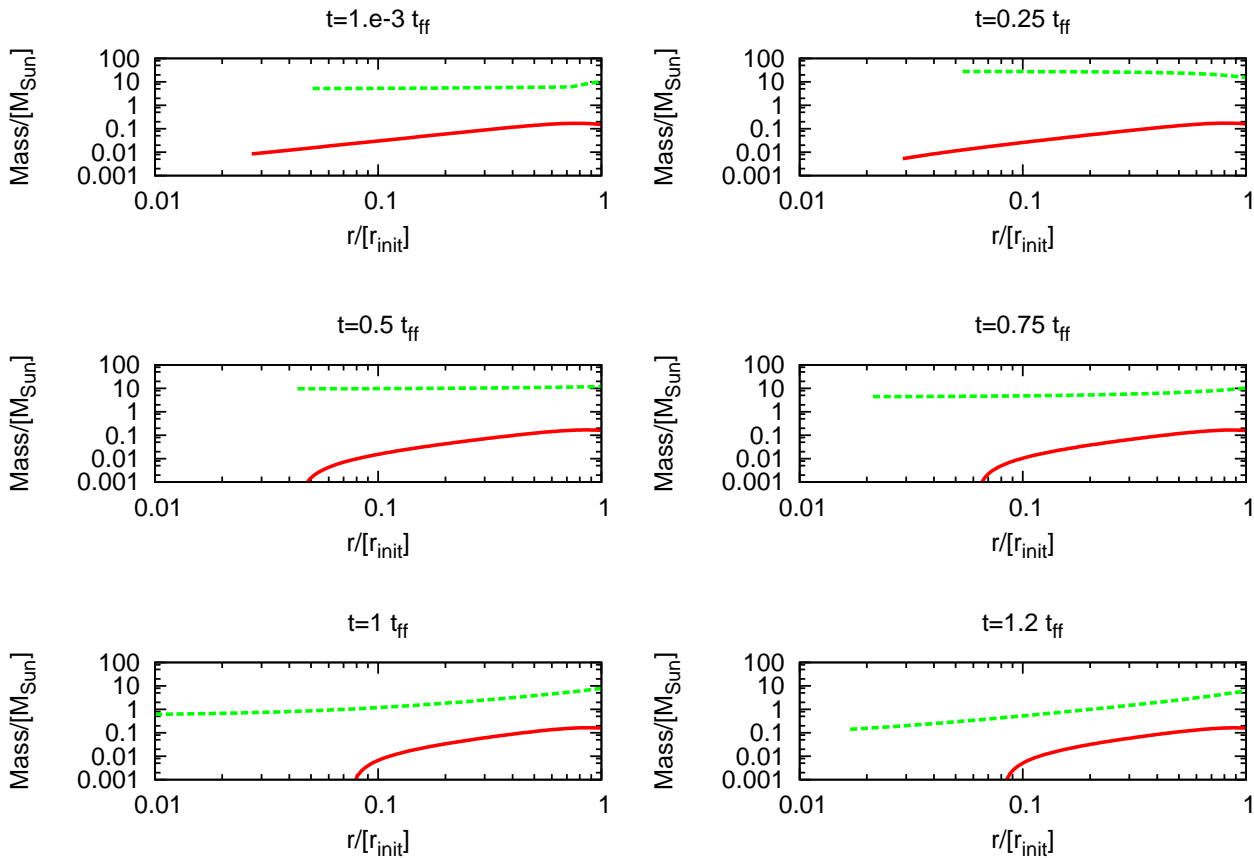}
  \caption{Shown here plots comparing the mass, $M(r)$, within a shell of radius $r$ (shown in red), and the Jeans mass, $M_{J}(r)$, for the respective shell (shown in green). Note that the plots are coeval with respect to the density and velocity characteristics shown previously in Fig. 9.}
\end{figure*}

Thus, shown in Fig. 10 is the mass, $M(r)$, within a shell of radius, $r$,(red characteristics), and the thermal Jeans mass, $M_{J}(r)$\footnote{$M_{J}(r)=(\pi/G)^{1.5}(a_{eff}^{3}(r)/\rho^{1/2}(r))$; $a_{eff}(r)$, being the effective sound-speed for a shell of radius $r$, and average density, $\rho(r)$.},(green characteristics), for these respective shells at different epochs. We note that the Jeans mass, $M_{J}(r)$, for a shell is progressively lowered as the polytrope contracts although, it is never overwhelmed by mass within that shell, $M(r)$. In other words, the compressional wave is not strong enough to assemble sufficient mass in a shell so that the Jeans stability criterion is never violated for that shell. The sound-speed in the calculation of $M_{J}(r)$ was replaced by the effective sound-speed to include the velocity dispersion induced by the inwardly propagating disturbance.

This plot is also supplemented by the one on the left-hand panel of Fig. B.1 that shows the BE mass, $M_{BE}^{core}(\xi\equiv r/R_{0})$, as a function of radius after about 1 free-fall time just before the assembled core rebounded. Like the Jeans mass, $M_{J}(r)$, the BE mass,  $M_{BE}^{core}(\xi)$, for a shell also remains higher than the corresponding mass of the shell. Consequently, the polytrope does not become self-gravitating. This plot in conjunction with those in Fig. 10 demonstrate that the Jeans criterion is essentially preserved at every radius within the polytrope, though Eqn. (8) leads to the converse conclusion. We may also add that the original result about the Super-Jeans nature of some starless cores by Sadavoy \emph{et al.}(2010), was also the consequence of assuming an isothermal gas. So although the Jeans stability criterion appears to be violated, in the strictest sense it is not, as we have demonstrated here.

To see if  further reduction of gas temperature close to the centre of the polytrope led to a collapse, we expanded on case 4 in the next case, listed 5 in Table1 by lowering the central temperature to 6 K while maintaining a warm annulus of gas as in case 4.  The polytrope in this case also evolved in a manner similar to the previous case and the average temperature of gas within the assembled core was now 6 K. A plot of its gas-density as a function of radius of the polytrope at different epochs over one free-fall time is shown on the upper panel of Fig. 11. Once again, it can be seen that having assembled a core at the centre of the polytrope, a rarefaction sets in and the central density is lowered. The velocity plots on the lower panel of this figure are therefore similar to those for case 4, plotted on the central-panel of Fig. 9.

In the final case, listed 6 in Table 1, the polytrope B was made strictly isothermal at 20 K, well below the critical temperature, $T_{0}$, defined by Eqn. (4). As in all the other cases discussed before this, the polytrope was perturbed by slightly increasing the external pressure, $P_{ext}$, that set off a compression-wave causing the initially gravitationally sub-critical polytrope to collapse and become singular. Evidently, thermal support in this case was insufficient to arrest the collapse. Also as demonstrated by the plot in the right-hand panel of Fig. B.1, the critical BE mass, $M_{BE}^{core}(\xi)$, is overwhelmed by the mass of the gas carried in by the compressional wave and so the polytrope becomes singular. Finally, shown in the left- and right-hand panel of Fig. 12 is respectively the radial density profile, and the radial component of gas-velocity within this polytrope. These plots are similar to those seen earlier for test cases 1, 2 and 3 where the respective polytropes also became singular; see also Figs. 3, 6, 7, 9 and 11.

\section{Discussion}
 Using the two polytropic models A and B  described in \S 2.4 above, we have attempted to reconcile some features of prestellar evolution. Results presented in \S 3 demonstrate the importance of initial conditions towards the formation and further evolution of a putative star-forming core and therefore determine its proclivity to spawn stars. Results from cases 1 and 2 demonstrate that the inherently unstable polytrope A, aided by an inwardly propagating compression-wave, was easily driven to collapse. The polytrope B used in the remaining 4 simulations on the other hand, presents a very interesting case when  perturbed similarly. The polytrope in cases 3 and 6 became singular as temperature of the gas within it was significantly lowered.

 On the contrary in test cases 4 and 5, where gas within the polytrope was maintained closer to the critical temperature, $T_{0}$, it exhibited intermittent phases of compression and expansion, a result that we believe, could help improve our understanding of the relatively long-lived prestellar cores. While the compression-wave within the polytrope in these latter cases largely remained subsonic, the radial inflow of gas assisted by this wave assembled a pressure supported core at the centre of the polytrope in each of these two cases. The core was perhaps only weakly bound.  Having reached the prestellar phase ($n\sim 10^{6}$ cm$^{-3}$), this core rebounded, as can be seen from the radial density and velocity plots shown in Figs. 9 and 11. 

Starless cores sometimes indeed, exhibit signs of inward motion as also reported by Myers \emph{et al.} (2000) for instance, and other references therein, for cores in the Perseus and Serpens molecular clouds. Other similar examples include the well-known starless core L1544 (e.g. Tafalla \emph{et al.} 1998), and a sample of those reported by Lee, Myers \& Tafalla  (1999, 2001),  Lee, Myers \& Plume (2004), and more recently by Schnee \emph{et al.} (2012). Inward-pointing velocity vectors within a core are generally considered to be positive evidence for gravitational boundedness. However, in view of the results from cases 4 and 5,  we suggest that while signatures of inward motion may be necessary to establish boundedness, they are certainly not sufficient.

Having imposed a more realistic temperature distribution defined by Eqn. (7) on gas within the model polytropes in the first five of our test simulations, they were rendered non-isothermal. The polytrope for case 6 though isothermal, was cold at a temperature lower than a third of the that required to maintain dynamical stability. The dynamical evolution of polytropes observed here shows that the temperature within their interiors, i.e., within the radius $r_{0}$, hardly affects their dynamical stability. For the compression-wave is set up in the outer regions of the polytrope and its strength depends on the temperature, $T_{out}$, of the gas in the annular region between the {\small ICM} and $r_{0}$. We will revisit this point in \S 4.1 where the possible implications for core-formation will be discussed. \\ \\
\emph{Other related work} We emphasise that the compression-wave observed within either polytropic model, A and B, is predicted by the analytically derived self-similar solutions for the Isothermal sphere. A full set of self-similar solutions derived by Whitworth \& Summers (1985), hereafter WS, show that those obtained earlier by Larson (1969), Penston (1969), Shu (1977), and Hunter (1977) belong to a larger family of solutions that  can be broadly divided into 3 bands, with those respectively in bands 0 and 1 being natural solutions. These latter solutions by WS showed that a compression-wave over-running a {\small SIS} or a {\small BES} could possibly trigger a runaway collapse in their respective configurations, as is indeed observed here in cases 1, 2, 3, and 6. The collapse of the polytrope observed in these latter test cases agrees with the classic outside-in scenario of the Larson-Penston solution, and unlike the inside-out Shu solution. Similarly, solutions by Tohline \emph{et al.} (1987) demonstrate that particularly strong fluctuations in the confining pressure could generate a compression-wave inducing the Isothermal sphere to collapse.

The more interesting results from cases  4 and 5 though, fall somewhere in-between the natural-solution bands of WS. Either the inward-moving wave in these two cases is not strong enough, or conversely, the initial polytrope, though unstable, is sufficiently warm so that this wave remains largely subsonic as can be seen in the velocity plots of Fig. 9. The formation of a pressure-supported core prevents further collapse of the interior regions. The plots in Figs. 9 and 11 offer an instructive comparison with those in Figs. 3, 6, 7, and 12. The agitated polytrope in these cases (4 and 5) is punctuated with phases in which the core mass appears to exceed its critical Bonnor-Ebert mass, $M_{BE}^{core}$, (see the plot in the lower panel of Fig. 9), and yet, it does not become singular. Although plots in Fig. 10 contradict this conclusion. These latter plots demonstrate that the inwardly propagating wave is not strong enough to  sufficiently raise the gas density, and therefore suitably lower the Jeans mass within the agitated polytrope so it could possibly be overwhelmed by the local gas mass. Consequently, there is no further collapse. 

\begin{figure}
  \vspace{5pt}
  \includegraphics[angle=270,width=8.5cm]{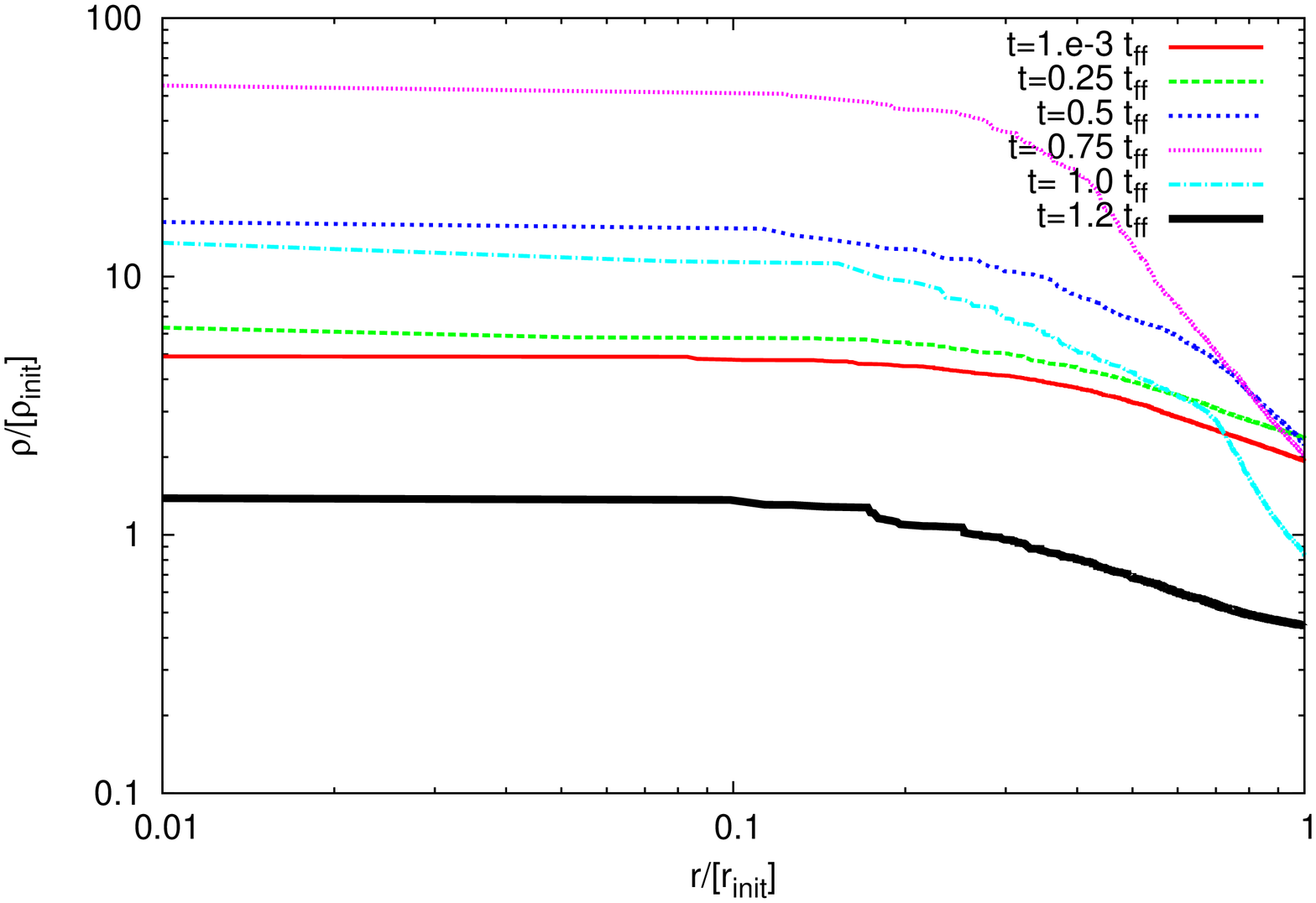}
  \includegraphics[angle=270,width=8.5cm]{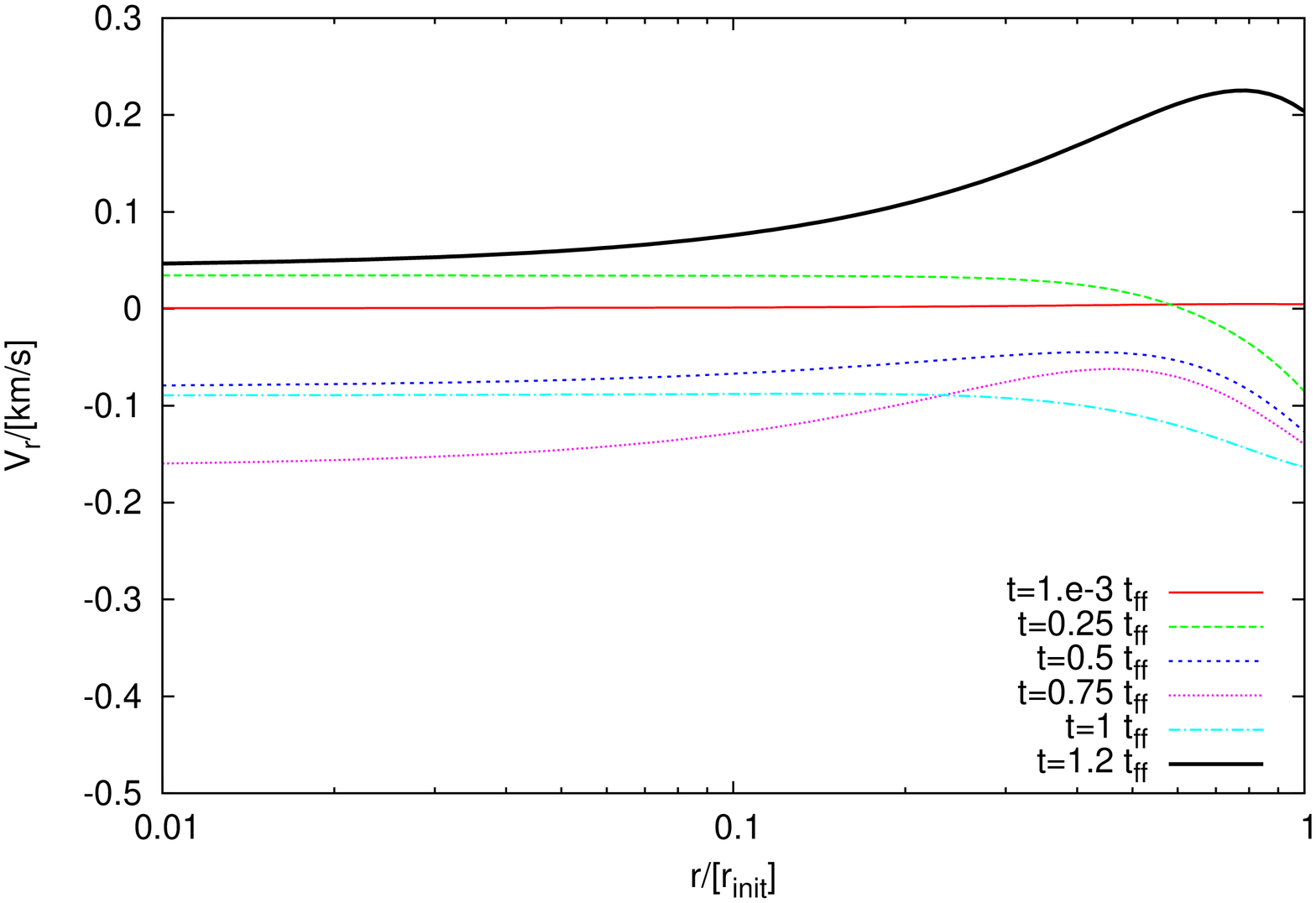}
  \caption{As in the upper-panel of Fig. 9, but for case 5. Shown here is the time-variation of the  radial profile for gas density within the polytrope on a logarithmic scale is shown. }
\end{figure}

This plot is further supported by another similar plot on the left-hand panel of Fig. B.1 which shows that like the Jeans mass, the BE mass in case 4 also remains significantly higher than the gas mass at that point so that formally there is no violation of the stability criterion. It is thus evident that cores could possibly be classified as "super-Jeans" under a set of simplifying assumptions like isothermality of gas. In reality though, they are far from isothermal and follow a temperature distribution similar to the one defined by Eqn. (7) above. The commonly observed non-thermal velocity fields within starless cores can also be reconciled by the results from cases 4 and 5.

 Polytropic models have also been tested by other authors. Hennebelle \emph{et al.} (2003), for instance, demonstrated that a Bonnor-Ebert sphere could be induced into collapse by sufficiently increasing the external pressure, $P_{ext}$. Gom{\' e}z \emph{et al.} (2007), more recently discussed one-dimensional models where core-rebound was observed in some cases. Importantly, however, the rebound in their simulations was a result of the gravitationally sub-critical nature of the core-envelope system. In other words, the criterion for gravitational stability was not violated in their simulations. Other authors, for instance,  Ballesteros-Paredes \emph{et. al.} (2003) have argued that cores forming in turbulent clouds are far from hydrostatic equilibrium and internal velocity dispersion is likely to delay their collapse. Turbulence though, is dissipative and loses energy rapidly on a timescale of the order of a free-fall time. It is therefore unlikely to support a core against gravity for too long. Furthermore, numerical simulations of turbulent cores show that, while some of the larger cores seem to produce filamentary structures, less massive cores produce protostellar accretion discs (e.g., Walch \emph{et al.} 2010). Thus, while turbulence may indeed delay a collapse, it may not succeed in arresting one. Also, there is little any corroborative evidence since most starless cores show subsonic, or at best, transonic velocity fields as has been discussed above.

It is clear that polytropic models are considerably successful in realising some dynamical properties of starless cores, though commenting  on the chemical properties of a typical starless core is difficult. CO and its isotopologues freeze out at high densities, i.e., at densities in excess of $10^{5}$ cm$^{-3}$ and temperatures less than $\sim$20 K (Di Francesco \emph{et al.} 2007). Depletion of these species would be expected in centrally condensed cores and such depletion has been observed in the starless cores L1498 and L1517B  (Tafalla \emph{et al.} 2006). Although on the one hand there is depletion of CO species, on the other, there is enhancement of deuterated species as CO depletion curtails the formation of DCO$^{+}$ via the coupling between H$_{2}$D$^{+}$ and CO (Crapsi \emph{et al.} 2005b). Importantly, the core assembled in test cases 4 and 5 
though ephemeral in nature, is suitably dense ($n\gtrsim 10^{6}$ cm$^{-3}$), and in fact, can perhaps sustain the associated gas-phase chemistry discussed above.

\begin{figure*}
  \vbox to 70mm{\vfil
  \includegraphics[angle=270,width=8.cm]{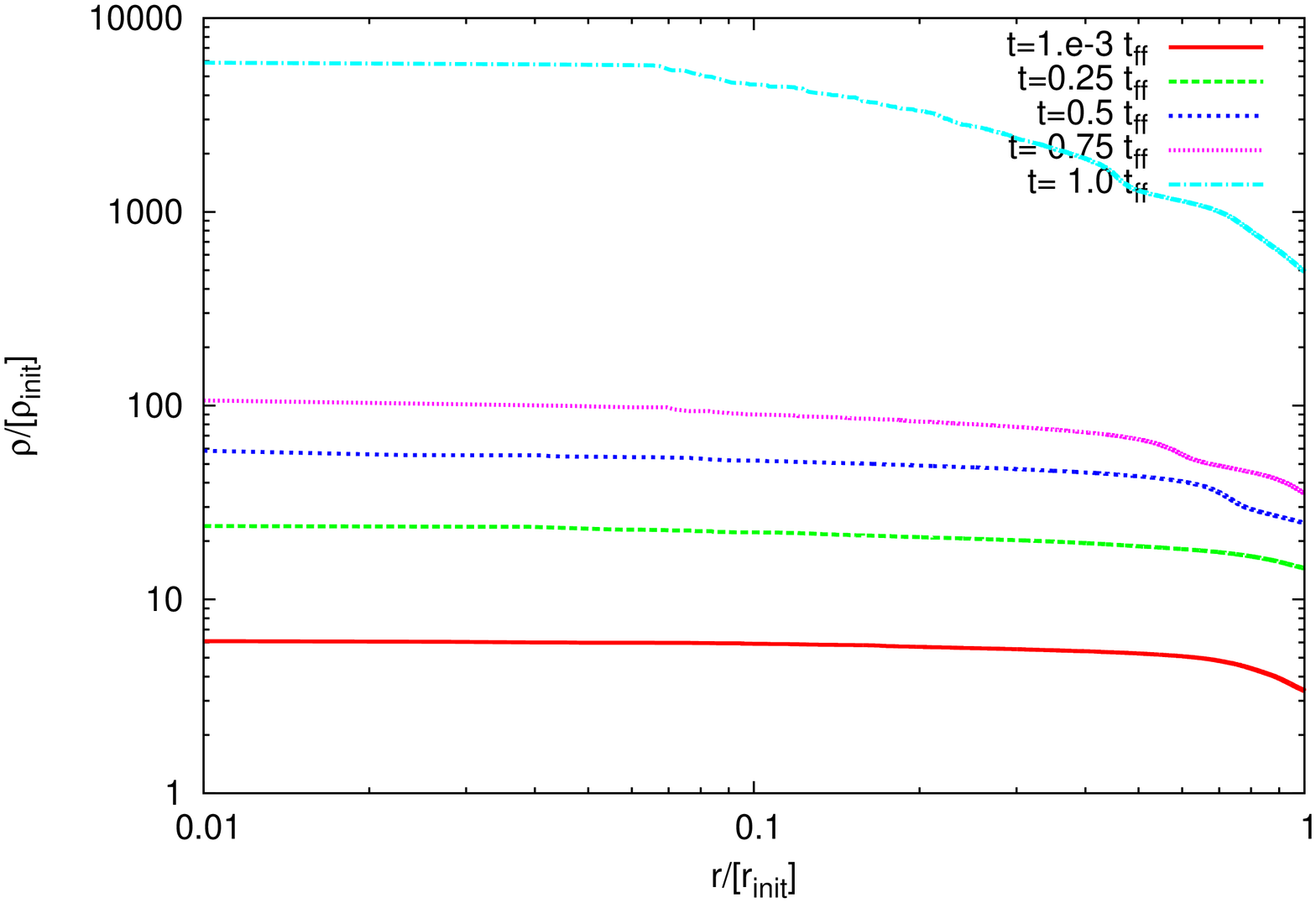}
  \includegraphics[angle=270,width=8.cm]{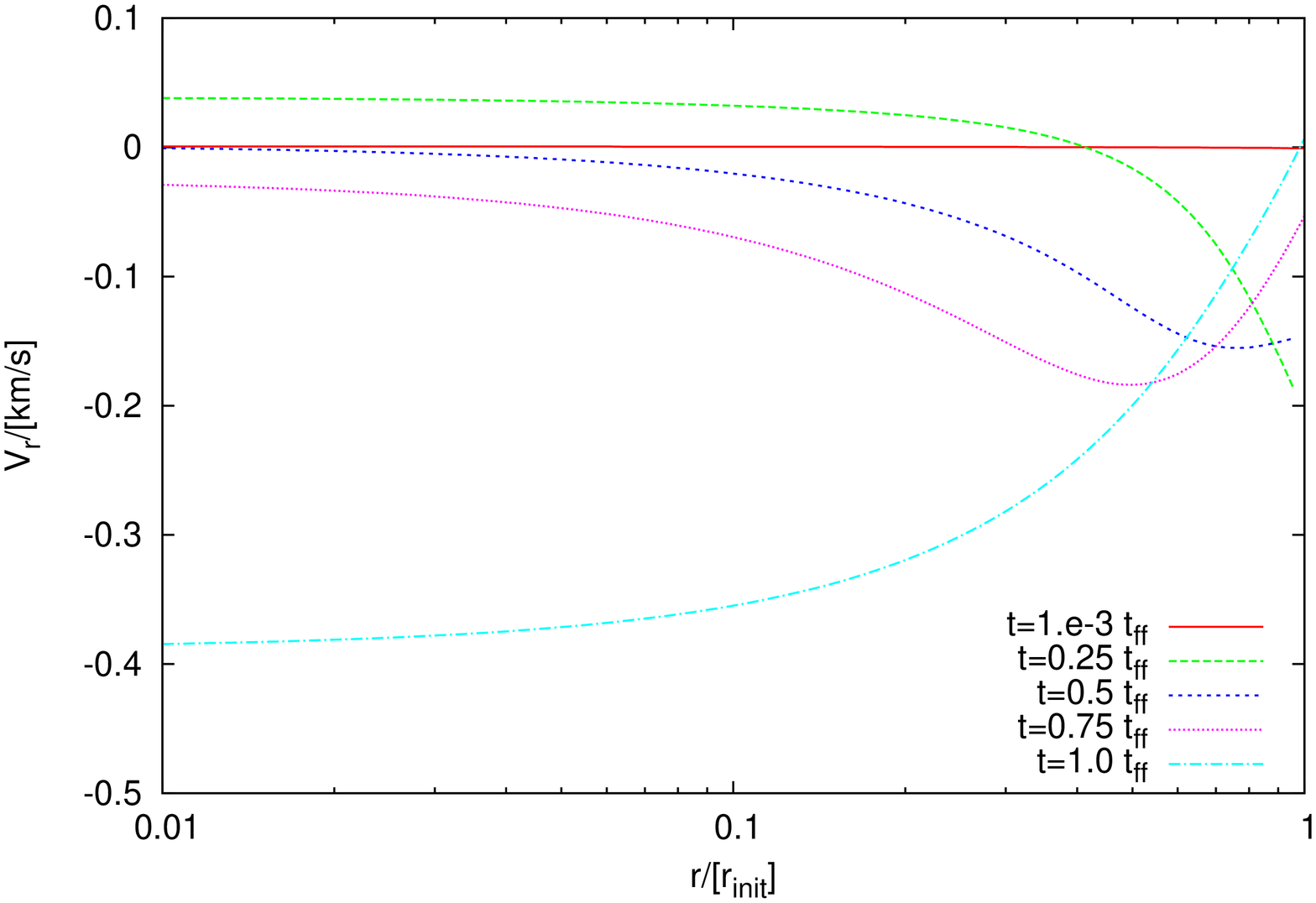}
  \caption{Coeval plots showing the radial profile of the gas-density and the radial component of the gas-velocity within the polytrope for case 6 have been plotted on respectively the left-, and right-hand panel. Either plots are similar to those seen previously for cases 1, 2 and 3 in Figs. 3, 6 and 7. The velocity plots show that as the compression-wave propagates inward, gas close to the centre, initially at rest, falls-in with ever increasing velocity as the polytrope becomes singular. }\vfil}
\end{figure*}

 Since there is not much variation of temperature during the expansion phase of the prestellar core, it is possible to reconcile apparent anomalies such as outward gas motions associated with chemical signatures of a typical dense starless core, or the converse. The starless core {\small L429}, for instance, has shown more red than blue asymmetry in observations of CS(2-1), indicating an outward motion (Lee \emph{et al.} 2001). Line profiles indicative of simultaneous infall and outward motions, i.e., radial oscillations, across the face of the well-known starless core Barnard 68 were noted by Lada \emph{et al.} (2003), and numerically modeled by Keto \emph{et al.} (2006). However,  oscillations demonstrated by these latter authors had a much smaller amplitude in comparison to those demonstrated here in test cases 4 and 5. Furthermore,  Lee \& Myers (2011) in their recent study of starless cores suggested that early in their life, these cores might remain static, then perhaps enter an oscillatory phase before finally contracting, having become sufficiently massive. 
\subsection{Implications for the formation and evolution of prestellar cores }
Supported by observations from two of our test models, 4 and 5, we infer that radial inflow of gas can readily assemble a thermally supported core.  Such flows are more likely to occur in regions dominated by hydrodynamic instabilities such as the Kelvin-Helmholtz, Rayleigh-Taylor, or the Thin-shell instability which readily generate filamentary clouds (e.g., Klessen \emph{et al.} 2005; Anathpindika 2009 a,b). See also Gong \& Ostriker (2009), where the formation of a core due to a supersonic, radial flow of gas was demonstrated.
Furthermore, a core assembled in this fashion, depending on the strength of compression and gas-temperature, could either become singular or remain dynamically unstable over several free-fall times. In view of this latter possibility, we suggest that the starlessness of some cores could be just a brief phase in the evolutionary cycle of these respective cores. 

The contrasting nature of results from the set of simulations discussed here show that temperature of the gas within the model polytrope is crucial towards its dynamical stability. Results from the present work as well as arguments presented by Tohline \emph{et al.}(1987) re-emphasise the importance of cooling in potential star-forming gas. In star-forming clouds a differential temperature distribution such as the one reported by Crapsi \emph{et al.} (2005), and exploited here could possibly arise due to variations in the efficiency of molecular cooling processes and local environment. Chemical processes and the environment responsible for cooling the gas during core formation therefore hold the key to the stability of that core.

\section{Conclusions}

Having accreted sufficient mass from its natal volume of gas a prestellar core 
becomes singular, spawning a young star within the collapsing core. Those cores in which  this collapse appears to have either halted or seemingly reversed into expansion despite a centrally peaked density distribution may remain starless. In the present study we have attempted to explain these features with the help of a polytropic model that utilises two varieties of the Isothermal sphere, viz., the singular and the pressure-supported type. Both polytropes were allowed to evolve under self-gravity after superposing a radially dependant temperature distribution.  The eventual fate of either polytropes appears to depend critically on the initial distribution of gas temperature within it.

A polytrope with cold gas is more likely to become singular when perturbed by a disturbance of suitable strength. A similar perturbation, on the other hand, can also set up oscillations within the polytrope when gas within it is relatively warm. In  the eventuality such as the later, each cycle of oscillation was observed to last for a few times 10$^{5}$ years. The oscillating polytrope acquired a centrally peaked density profile even as density in the outer regions  followed a relatively shallow power-law of the type, $\rho(r)\propto r^{-k}$,  similar to that observed in typical starless cores. We also report that in each of these contraction phases, mass within the central dense region, referred to as a core in this study, appears to exceed its critical mass for dynamical stability though it does not formally violate the stability criterion. Supported by this conclusion we suggest that the attribute "Super-Jeans" is subjective and could be potentially mis-leading. The assembled core survived for a few 10$^{4}$ yrs before rebouncing during a rarefaction, only to be reassembled in the following cycle of compression. We therefore suggest, starlessness is probably only a short, temporary phase in the lifespan of a pressure-supported core. Our simulations also offer a possible explanation for the origin of a non-thermal velocity field within a starless core and show that an inwardly directed velocity field is a necessary but not sufficient condition to establish its boundedness. 

 However, it would be useful to repeat the present set of numerical tests with an Eulerian grid-based algorithm to further confirm the conclusions arrived at in this work.

Without invoking the magnetic field or an external turbulent velocity field, the models discussed here reconcile the relatively long lifetimes of some cores. We have also demonstrated that prestellar cores can be assembled by a radial inflow of gas in star-forming clouds which is particularly likely in filamentary clouds. We also suggest that while inward pointing velocity vectors necessarily indicate inward gas motion, it is insufficient to establish the gravitational boundedness of a core with any degree of certainty. Extensive mapping of velocity fields within cores is essential to establish if indeed their directions can be used as unambiguous prognoses of the prestellar phase. It must be warned that although a polytropic model such as the one discussed in the present study appears commensurate with the dynamical properties of starless cores, it is not completely clear how in star-forming clouds such an assembly would be realised. The study of core formation and the effects of cooling processes on the local environment therefore deserve further attention.

\section*{Acknowledgements}

The authors are grateful to an anonymous referee for suggestions that helped improve the original manuscript and make results clearer. Sumedh Anathpindika (SA) also acknowledges support by a post-doctoral fellowship of the Department of Science \& Technology of the government of India. SA thanks David Hubber for sharing the numerical code {\small SEREN}, while the rendered density plots were prepared using the publicly available graphics package, {\small SPLASH} prepared by Daniel Price. Part of this paper was originally completed at the University College London through a generous visitor's grant which SA acknowledges. James Di' Francesco acknowledges the generous hospitality received at the Indian Institute of Astrophysics during his visit that made this work possible.

\appendix 
\section{The critically stable Isothermal sphere} 
This simple test was performed to demonstrate that the {\small SPH} algorithm could in fact reproduce the analytically expected singular nature of the Isothermal sphere having radius, $\xi_{B}=\xi_{crit}=6.45$.
As in the main body of the paper, we first assemble and settle the Isothermal sphere, now for $\xi_{B}\sim 6.45$. The settled polytrope was then rescaled so that it had a radius, $r_{init}$ = 0.1 pc, and mass, $M_{init}$ = 1 M$_{\odot}$. Gas within the polytrope was held at a uniform temperature of 5 K (lower than the critical temperature, $T_{0}$, defined by Eqn. (4), which in this case is ~14 K), surrounded by a jacket of particles mimicking the intercloud medium, maintained at 10 K. Without much surprise, the polytrope begins to collapse as can be seen from the radial density distribution plotted in Fig. A1 at two epochs.  

\begin{figure}
\vspace{2pc}
 \includegraphics[angle=270,width=8.cm]{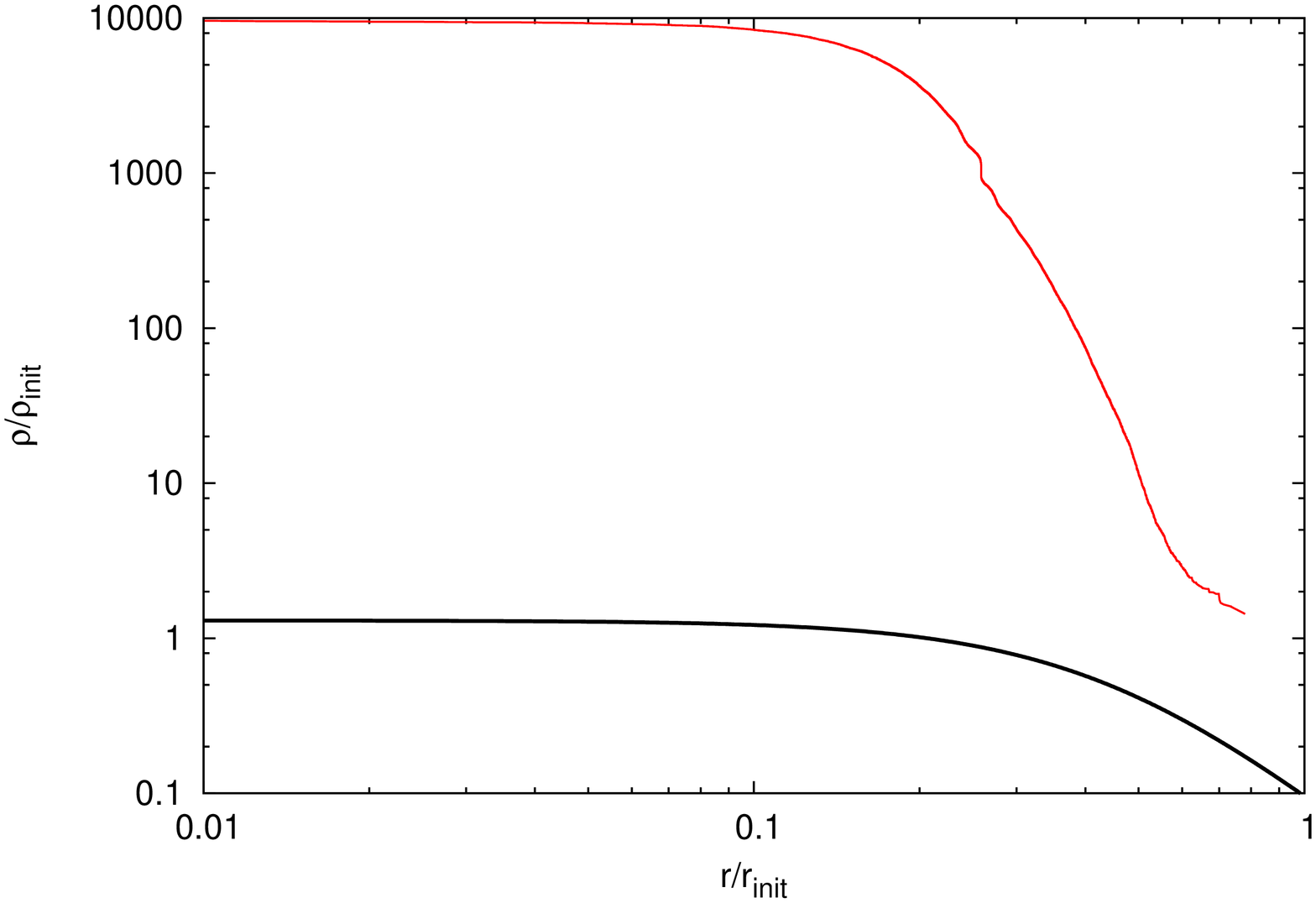}
\caption{Radial density profile of the Isothermal sphere with $\xi_{B}\sim6.45$. The characteristics have been plotted at $t\sim 0$ (black), and $t\sim 0.3 t_{ff}$ (red); $t_{ff}\sim 1$ Myr is the free-fall time of the Isothermal sphere.}
\label{appenfig}
\end{figure}

\section{Can a starless core be Super-Jeans ?}
The answer to this question is unlikely to be affirmative though some cores, such as the one assembled in this work,may appear to remain starless despite exceeding their thermal Jeans and/or the Bonnor-Ebert mass; they do not become singular. Plots in Figs. 10 and B.1 demonstrate that the compressional wave in the original polytrope in case 4 fails to assemble enough gas so that neither the Jeans criterion, nor the BE criterion for stability is violated and the respective polytrope fails to become singular. Instead it performed large amplitude radial oscillations and the assembled core rebounded. Under the assumption of isothermality though, the core appears to have violated the stability criteria as shown in \S 3.2. This conclusion, however, is contradicted by a little more careful analysis.

\begin{figure*}
  \hspace{1.cm}
  \begin{minipage}{\linewidth}
  \centering 
   \mbox{\includegraphics[angle=270,width=8.5cm]{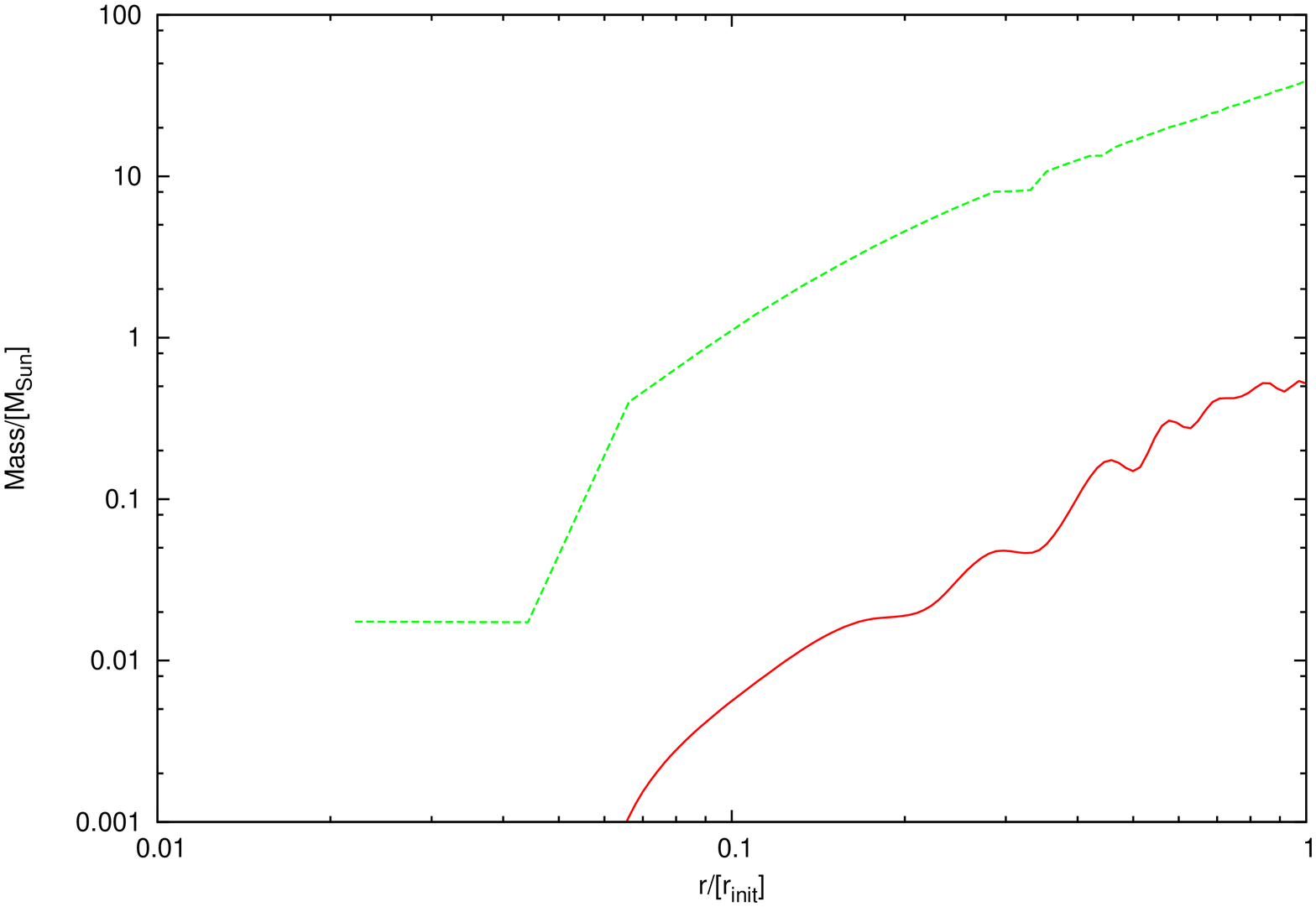}
        \includegraphics[angle=270,width=8.5cm]{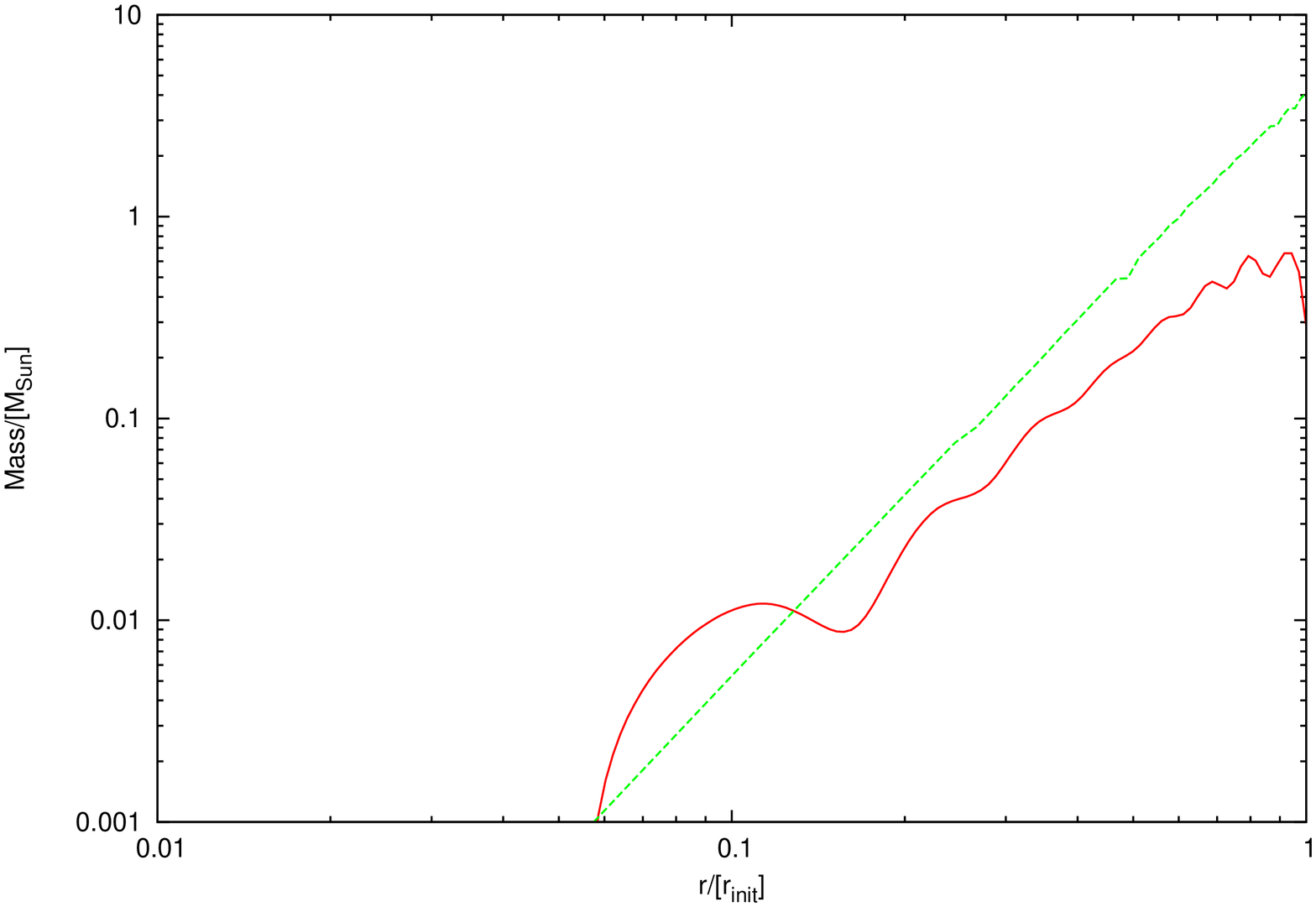}}
  \caption{Shown in these plots is the Bonnor-Ebert({\small BE}) mass (green characteristic) as a function of the radial distance within the polytrope. Also shown is the corresponding mass, $M(r)$(red characteristic), at that radius. The plots in the left and the right-hand panel correspond respectively, to cases 4 and 6 after a free-fall time($\sim$ 0.3 Myr). The core mass in case 4 fails to overcome its {\small BE} mass so that the polytrope fails to collapse. On the contrary, the polytrope in case 6 becomes singular as mass close to the centre of the polytrope exceeds the corresponding {\small BE} mass. }
 \end{minipage}
\end{figure*}

\bsp

\label{lastpage}


\begin{thebibliography}{99}

\bibitem[\protect\citeauthoryear{Anathpindika}{2009}]{A3}Anathpindika, S., 2009a, A\&A, 504, 437
\bibitem[\protect\citeauthoryear{Anathpindika}{2009}]{A4}Anathpindika, S., 2009b, A\&A, 504, 451
\bibitem[\protect\citeauthoryear{Andre}{2007}]{A2}Andr{\' e}, Ph., Belloche, A., Motte, F \& Peretto, N., 2007, A\&A, 472, 519
\bibitem[\protect\citeauthoryear{bALLESTEROS}{2003}]{b2} Ballesteros-Paredes, J., Klessen, R \& V{\' a}zquez-Semadeni, E., 2003, ApJ, 592, 188
\bibitem[\protect\citeauthoryear{Bonnor}{1956}]{b1} Bonnor, W., 1956, MNRAS, 116, 351
\bibitem[\protect\citeauthoryear{Barnes}{1986}]{b2} Barnes, J \& Hut, P., 1986, Nature, 324, 446
\bibitem[\protect\citeauthoryear{Bate}{1995}]{b3}Bate, M., Bonnell, I \& Price, N., 1995, MNRAS, 277, 362
\bibitem[\protect\citeauthoryear{Bate}{1997}]{b4}Bate, M \& Burkert, A., 1997, MNRAS, 288, 1060
\bibitem[\protect\citeauthoryear{Belloche}{2011}]{b4}Belloche, A., Parise, B., Schuller, F., Andr{\' e}, Ph., Bontemps, S \& Menten, K., 2011, \emph{A\&A accepted}, astroph 1106.5064
\bibitem[\protect\citeauthoryear{Chandrasekar}{1939}]{C1} Chandrasekar, S., 1939, \emph{An Introduction to the study of stellar structure}, Dover Pub. Inc.
\bibitem[\protect\citeauthoryear{Crapsi et al.}{2005}]{C2} Crapsi, A., De Vries, C., Huard, J., Lee, J., Myers, P., Ridge, N. \emph{et al.}, 2005a, A\&A, 439, 1023
\bibitem[\protect\citeauthoryear{Crapsi et al.}{2005}]{C3} Crapsi, A., Caselli, P., Walmsley, M., Myers, P. C., Tafalla, M., Lee, C \& Bourke, T., 2005b, ApJ, 619, 379
\bibitem[\protect\citeauthoryear{Crapsi et al.}{2007}]{C4} Crapsi, A., Caselli, P., Walmsley, M \& Tafalla, M., 2007, A\&A, 470, 221
\bibitem[\protect\citeauthoryear{Ciolek}{2001}]{C5}Ciolek, G. E. \& Basu, S., 2001, ApJ, 547, 272
\bibitem[\protect\citeauthoryear{Francesco}{2007}]{D1}Di Francesco, J., Evans, N. J II., Caselli, P., Myers, P. C., Shirley, Y., Aikawa, Y \& Tafalla, M., 2007, \emph{Protostars and Planets V}, 951, 17
\bibitem[\protect\citeauthoryear{Francesco}{2008}]{D2}Di Francesco, J., Johnstone, D., Kirk, H., MacKenzie, T \& Ledwosinska, E., 2008, ApJSS, 175, 277
\bibitem[\protect\citeauthoryear{Ebert}{1955}]{E1} Ebert, R., 1955, ZA, 36, 222
\bibitem[\protect\citeauthoryear{Evans}{2001}]{E2} Evans, N. J. II, Rawlings, J. M., Shirley, Y. L \& Mundy, L. G., 2001, ApJ, 557, 193
\bibitem[\protect\citeauthoryear{Frau}{2010}]{F1} Frau, P., Girart, J. M., Beltr{\' a}n, T. M., Morata, O \emph{et al.}, 2010, ApJ, 723, 1665
\bibitem[\protect\citeauthoryear{Galvan}{2007}]{G3} Galv{\' a}n-Madrid, R., V{\' a}zquez-Semadeni, E., Kim, J \& Ballesteros-Paredes, J., 2007, ApJ, 670, 480
\bibitem[\protect\citeauthoryear{Goldsmith}{2001}]{G2} Goldsmith, P F., 2001, ApJ, 557, 736
\bibitem[\protect\citeauthoryear{Gomez}{2007}]{G1} Gom{\' e}z, G., V{\' a}zquez-Semadeni, E., Shadmehri, M \&  Ballesteros-Paredes, J., 2007, ApJ, 669, 1042
\bibitem[\protect\citeauthoryear{Gong}{2009}]{G4} Gong, H \& Ostriker, E., 2009, ApJ, 699, 230
\bibitem[\protect\citeauthoryear{Harvey}{2003}]{H4} Harvey, D W. A., Wilner, D. J., Lada, C. J., Myers, P. C. \& Aves, J. F., 2002, ApJ, 598, 1112
\bibitem[\protect\citeauthoryear{Harvey}{2007}]{H5} Hatchell, J., Fuller, G. A., Richer, J. S., Harries, T. J \& Ladd, E. F., 2007, A\&A, 468, 1009
\bibitem[\protect\citeauthoryear{Hennebelle}{2003}]{H1}Hennebelle, P., Whitworth, A., Gladwin, P \& Andr{\' e}., P., 2003, MNRAS, 340, 870
\bibitem[\protect\citeauthoryear{Hubber et al.}{2011}]{H2} Hubber, D., Batty, C., McLeod, A \& Whitworth,A., 2011, A\&A accepted, astroph 1102.0721 
\bibitem[\protect\citeauthoryear{Hunter}{1977}]{H3}Hunter, C., 1977, ApJ, 218, 834
\bibitem[\protect\citeauthoryear{Jijina}{1999}]{J1}Jijina, J., Myers, P \& Adams, F., 1999, ApJSS, 125, 161
\bibitem[\protect\citeauthoryear{Kirk}{2007}]{K1}Kirk, J., Ward-Thompson, D \& Andr{\' e}, P., 2007,  MNRAS, 375, 843
\bibitem[\protect\citeauthoryear{Keto}{2006}]{K2} Keto, E., Broderick, A. E., Lada, C. J. \& Narayan, R., 2006, ApJ, 652, 1366
\bibitem[\protect\citeauthoryear{Klessen}{2005}]{K3}Klessen, R., Ballesteros-Paredes, J., V{\' a}zquez-Semadeni, E \& Dur{\' a}n-Rojas, C., 2005, ApJ, 620, 786
\bibitem[\protect\citeauthoryear{Larson}{1969}]{L5} Larson, R. B., 1969, MNRAS, 145, 271
\bibitem[\protect\citeauthoryear{Lada}{2003}]{L4} Lada, C. J., Bergin, E. A., Alves, J. F. \& Huard, T. L., 2003, ApJ, 586, 286
\bibitem[\protect\citeauthoryear{Lee}{1999}]{L1}Lee, C. W., Myers, P. C \& Tafalla, M., 1999, ApJ, 526, 788
\bibitem[\protect\citeauthoryear{Lee}{2001}]{L2}Lee, C. W., Myers, P. C \& Tafalla, M., 2001, ApJSS, 136, 703
\bibitem[\protect\citeauthoryear{Lee}{2004}]{L3}Lee, C. W., Myers, P. C \& Plume, R., ApJSS, 153, 523
\bibitem[\protect\citeauthoryear{Lee}{2011}]{L4}Lee, C. W \& Myers, P. C., 2011, ApJ, 734, 60
\bibitem[\protect\citeauthoryear{Monaghan}{2005}]{M1} Monaghan, J., 2005, \emph{Rep. Prog. in Phys.}, 68, 1703
\bibitem[\protect\citeauthoryear{Monaghan}{2000}]{M3}Myers, P. C., Evans, N. J. II \& Onishi, N., 2000, in Protostars \& Planets IV, ed. Mannings, \emph{et. al.} (Tuscon university, Arizona), 217
\bibitem[\protect\citeauthoryear{Pavlovski}{2002}]{P1}Pavlovski, G., Smith, M. D., Mac Low, M. M. \& Rosen, A., 2002, MNRAS, 337, 477
\bibitem[\protect\citeauthoryear{Penston}{1969}]{L4} Penston, M. V., 1969, MNRAS, 144, 425
\bibitem[\protect\citeauthoryear{Redman}{2005}]{R1} Redman, M.P., Keto, E \& Rawlings, J.M.C., 2006, MNRAS, 370, L1
\bibitem[\protect\citeauthoryear{Sadavoy et al.}{2010}]{S1}Sadavoy, S., Di Francesco, J., Botemps, S., Megeath, S., Rebull, L., Allgaier, E. \emph{et al.}, 2010a, ApJ, 710, 1247
\bibitem[\protect\citeauthoryear{Sadavoy}{2010}]{S2}Sadavoy, S., Di Francesco, J \& Johnstone, D., 2010b, ApJ, 718, L32
\bibitem[\protect\citeauthoryear{Schnee}{2012}]{S4} Schnee, S., Sadavoy, S., Di' Francesco, J., Johnstone, D \& Wei, L., 2012, ApJ, 755, art. id. 178
\bibitem[\protect\citeauthoryear{Shu}{1977}]{S3} Shu, F., 1977, ApJ, 214, 488
\bibitem[\protect\citeauthoryear{Simpson}{2011}]{S4} Simpson, R. J., Johnstone, D., Nutter, D., Ward-Thompson, D \& Whitworth, A., 2011, \emph{accepted by MNRAS}, astroph 1106.1885
\bibitem[\protect\citeauthoryear{Stamattelos}{2007}]{S4} Stamatellos, D., Whitworth, A. P., Bisbas, T \& Goodwin, S., 2007, A\&A, 475, 37
\bibitem[\protect\citeauthoryear{Tafalla}{1998}]{T1} Tafalla, M., Mardones, D., Myers, P. C., Caselli, P., Bachiller, R. \emph{et al.}, 1998, ApJ, 504, 900
\bibitem[\protect\citeauthoryear{Tafalla}{2004}]{T3} Tafalla, M., Myers, P. C., Caselli, P \& Walmsley, C. M., 2004, A\&A, 416, 191
\bibitem[\protect\citeauthoryear{Tafalla}{2004}]{T4} Tafalla, M \& Santiago, J., 2004, A\&A, 414, L53
\bibitem[\protect\citeauthoryear{Tafalla}{2006}]{T2} Tafalla, M., Santiago-Garc{\' i}a, J., Myers, P. C., Caselli, P., Walmsley, C. M. \& Crapsi, A., 2006, A\&A, 455, 577
\bibitem[\protect\citeauthoryear{Tohline}{1992}]{b45}Tohline, J. E, Bodenheimer, P. H. \& Christodoulou, D. M., 1987, ApJ, 322, 787
\bibitem[\protect\citeauthoryear{Vazquez}{2005}]{V1}V{\' a}zquez-Semadeni, E., Kim, J., Shadmehri, M \& Ballesteros-Paredes, J., 2005, ApJ, 618, 344
\bibitem[\protect\citeauthoryear{Ward-Thompson}{1994}]{W1}Ward-Thompson, D., Scott, P. F., Hills, R. E \& Andr{\' e}, P., 1994, MNRAS, 268, 276
\bibitem[\protect\citeauthoryear{Ward-Thompson}{2007}]{W3}Ward-Thompson, D., Di Francesco, J., Hatchell, J., Hogerheijde, M. R., Nutter, D., Bastien, P \emph{et al.}, 2007, PASP, 119, 855
\bibitem[\protect\citeauthoryear{Ward-Thompson}{2010}]{W2}Ward-Thompson, D., Kirk, J., Andr{\' e}., P., Saracero, P. \emph{et al.}, 2010, A\&A, 518, L92
\bibitem[\protect\citeauthoryear{Walch}{2010}]{W3} Walch, S., Naab, T., Whitworth, A., Burkert, A \& Gritschneder, M., 2010, MNRAS, 402, 2253
\bibitem[\protect\citeauthoryear{Whitworth}{1985}]{W2}Whitworth, A \& Summers, D., 1985, MNRAS, 214, 1 (WS)
\end{thebibliography}
\end{document}